\documentclass[%
  reprint,
  superscriptaddress,
   amsmath,amssymb,
   aps,
   pre,
]{revtex4-2}

\usepackage{graphicx}

\usepackage{dcolumn}
\usepackage{bm}
\usepackage{hyperref}
\usepackage[mathlines]{lineno}
\usepackage{braket}
\usepackage{float}
\usepackage{comment}
\usepackage{mathtools}
\DeclarePairedDelimiter{\abs}{\lvert}{\rvert}

\begin{document}


\title{Pink-noise dynamics in an evolutionary game on a regular graph}

\author{Yuki Sakamoto}
\affiliation{Department of Physics, The University of Tokyo, 7-3-1 Hongo, Bunkyo-ku, Tokyo 113-0033, Japan}
\author{Masahito Ueda}%
\affiliation{Department of Physics, The University of Tokyo, 7-3-1 Hongo, Bunkyo-ku, Tokyo 113-0033, Japan}
\affiliation{Institute for Physics of Intelligence, Graduate School of Science, The University of Tokyo, Bunkyo-ku, Tokyo 113-0033, Japan}
\affiliation{RIKEN Center for Emergent Matter Science (CEMS), Wako, Saitama 351-0198, Japan}

\date{\today}
\begin{abstract}
  We consider a multiplayer prisoner's dilemma game on a square lattice and regular graphs based on the pairwise-Fermi update rule,
  and obtain heatmaps of the fraction of cooperators and the correlation of neighboring pairs.
  In the heatmap, there is a mixed region where cooperators and defectors coexist,
  and in the mixed region the correlation between neighbors is enhanced.
  Moreover, we observe pink-noise behavior in the mixed region, where the power spectrum can be fitted by a power-law function of frequency.
  We also find that the pink-noise behavior can be reproduced in a simple random-walk model.
  In particular, we propose a modified random-walk model which can reproduce not only the pink-noise behavior but also the deviation from it observed in a low-frequency region.
\end{abstract}

\maketitle


\section{\label{sec:introduction}Introduction\protect}

The history of modern game theory dates back to von Neumann, who published a seminal paper on mixed-strategy equilibria in the two-person zero-sum game \cite{vonNeumann:1928:TGG}\cite{vonNeumann:1959}.
Later his idea was applied to economics \cite{gibbons1992game}.
von Neumann also collaborated with Oskar Morgenstern to write a book entitled ``Theory of Games and Economic Behavior'' \cite{morgenstern1953theory}.

Modern game theory has been applied to various fields such as politics \cite{Downs:1957}, psychology \cite{camerer2011behavioral} and biology \cite{maynard1973logic}.
Two key concepts of game theory are Nash equilibria and Pareto optimality.
John Forbes Nash discussed equilibria in a strategic-form noncooperative game and proved the existence of equilibria, which are referred to as Nash equilibria \cite{nash1950equilibrium}.
Vilfredo Frederico Damaso Pareto created a concept known as the Pareto optimality, where payoffs are most efficiently distributed so that nobody cannot increase his/her payoff without making another player worse off \cite{osborne1994course}.

In particular, some game theoretical-models are considered to explain development of cooperative relation even in social viscosity \cite{ito2018scaling} in a large population \cite{wang2015universal}.
In recent years, several mechanisms have been found to construct sustainable cooperation by giving players mutual advantages and keeping them from undesirable risk avoidance.
One mechanism is a network structure \cite{nowak1992evolutionary}.
When we consider a large well-mixed population where individuals equally match with others, iterated games with dilemma may cause natural selection;
without any mechanism for cooperation, cooperators are likely to be extinct.
On the other hand, when players are not mixed well and different individuals tend to interact with different specific others, such a relation can be expressed by a graph \cite{wasserman1994social};
each individual is represented by a node, and each interaction between individuals is represented by an edge.
Considering a case where iterated games with dilemma on a network, we observe mutual cooperation, unlike a well-mixed case \cite{nowak1992evolutionary}.
Other mechanisms are also known to enhance evolutionary cooperation,
such as kin selection, direct reciprocity, indirect reciprocity and group selection \cite{nowak2006five}.

In general, games can be classified into strategic-form games and extensive-form games \cite{osborne1994course}.
A strategic-form game is specified by three components, i.e., players, strategies and payoffs.
An $n$-player strategic-form game is given by a set $M=\set{1,\cdots,n}$ of $n$ players, a set $\set{S^i}_{i\in M}$ of strategies available for players and a set
$\set{f^i}_{i\in M}$ of payoff functions for players:
\begin{eqnarray}
  \label{eq:components_games}
  (M,\set{S^i}_{i\in M},\set{f^i}_{i\in M}).
\end{eqnarray}
Here the payoff function $f^i:S^1\times\cdots\times S^n\rightarrow\mathbb{R}$ is a real-valued function of joint strategies and gives a payoff to player $i$ according to strategies $(s^1,\cdots,s^n)\in S^1\times\cdots\times S^n$ taken by players.
We express the direct product of strategies of $n-1$ players with one player $i$ excluded as $S^{-i}\equiv S^1\times\cdots S^{i-1}\times S^{i+1}\times\cdots\times S^n$.
In addition, a component of the direct product is denoted by $s^{-i}\equiv (s^1,\cdots,s^{i-1},s^{i+1},\cdots,s^n)$.

In a strategic-form game, each player chooses a strategy from his/her set of strategies.
When players have a finite number of available strategies, a strategic-form game can simply be expressed by a payoff matrix.
By way of illustration, let us consider a situation in which there are two players $1, 2$,
and each player $i(=1,2)$ has a strategy set $S^i=\set{s^i_1,\cdots,s^i_{m^i}}$ of $m^i$ strategies.
When players $1$ and $2$ choose strategies $s^1_i$ and $s^2_j$, respectively, players $1$ and $2$ gain payoffs $a^1_{ij}=f^1(s^1_i,s^2_j)$ and $a^2_{ij}=f^2(s^1_i,s^2_j)$, respectively.
Then the game can be represented by assigning the joint payoffs $(a^1_{ij}, a^2_{ij})$ to the $(i,j)$-th component of a matrix.
This matrix is referred to as the payoff matrix, and the matrix $A$ in TABLE \ref{table:two_by_two} is the payoff matrix for \(m^i=2\) and \(i=1,2\).

\begin{table}[H]
\caption{$2\times 2$ payoff matrix $A$ for a strategic-form game.}
\label{table:two_by_two}
  \centering
  \begin{tabular}{c|cc}
    $1$\textbackslash $2$ & $s^2_1$ & $s^2_2$\\ \hline
    $s^1_1$ & $(a^1_{11},a^2_{11})$ & $(a^1_{12},a^2_{12})$\\
    $s^1_2$ & $(a^1_{21},a^2_{21})$ & $(a^1_{22},a^2_{22})$
  \end{tabular}
\end{table}

In the above-mentioned strategic-form game, all players simultaneously choose their strategies.
We can also consider a game where players determine their strategies at different times.
Such a game is referred to as an extensive-form game.
Examples of strategic-form games are the prisoner's dilemma game \cite{axelrod1980effective}, the stag-hunt game \cite{skyrms2004stag} and the battle of sexes \cite{cooper1989communication}.
Examples of extensive-form games include chess and poker \cite{binmore2007playing}.

Here we briefly describe strategic-form games which constitute the main theme of this paper.
Strategic-form games can be classified into games with perfect information and games with imperfect information \cite{osborne1994course}.
They can also be classified into cooperative games \cite{davis1965kernel} and noncooperative games \cite{nash1951non}.
In a game with perfect information, the three components described in Eq. (\ref{eq:components_games}) are all common knowledge to all players.
In contrast, in a game with imperfect information, not all the three components in Eq. (\ref{eq:components_games}) are common knowledge \cite{osborne1994course}.
Meanwhile, in a noncooperative game, each player independently chooses his strategy without prior negotiation.
In contrast, in a cooperative game, players are allowed to negotiate with others before choosing their strategies.

In this paper, we focus on a noncooperative game with imperfect information.
In particular, we discuss evolutionary games, where multiple players interact with each other and change their strategies in an attempt to increase their payoffs over time.
Evolutionary game theory was initiated by John Maynard Smith and George Robert Price \cite{maynard1973logic}, where the concept of Nash equilibria was developed into a new idea of an evolutionarily stable strategy \cite{dugatkin2000game}\cite{newton2018evolutionary}.
This idea of evolutionary games has widely been applied to biology \cite{anderson1982coevolution} and genetics \cite{buss1989sex}.

We examine the dynamics of the fraction of cooperators in an iterated prisoner's dilemma game and discuss the power spectrum of the fraction of cooperators.
We find that the obtained power spectrum behave as a pink noise, which means that the spectrum is proportional to \(S(f)\propto f^{-\alpha}\),
where \(\alpha\) is a positive constant.
The pink-noise behavior is widely studied and is a ubiquitous phenomenon in physics.
It is of interest that a similar pink-noise behavior can also emerge in the game dynamics.

We demonstrate that a simple random-walk model reproduce the pink-noise behavior of the fraction of cooperators in our game model.
We show that the dynamics of the fraction of cooperators in a game-theoretical model can be reproduced by a much simpler random-walk model.
We find that the power exponent of the spectrum can also be reproduced by the random-walk model, and that
in some cases the power spectrum of the fraction of cooperators deviates from the pink-noise curve in low-frequency region.
In such cases, we show that the deviation of the spectrum can be reproduced by a modified random-walk model.

This paper is organized as follows.
In Sec. \ref{sec:evolutionary_game}, we briefly introduce fundamental ideas of game theory and give the setting of the iterated game based on the prisoner's dilemma game.
In Sec. \ref{sec:ensemble-averaged_square} and Sec. \ref{sec:ensemble-averaged_regular}, we describe the numerical results of ensemble-averaged quantities in the games on a square lattice and regular graphs, respectively.
In Sec. \ref{sec:single}, we numerically examine dynamical trajectories of the fraction of cooperators, and we find that their spectra exhibit pink-nose behavior.
In Sec. \ref{sec:discussion}, we discuss how pink-noise behavior is reproduced in terms of random-walk models.
In Sec. \ref{sec:conclusion}, we describe a summary and an outlook.

\section{\label{sec:evolutionary_game}Evolutionary Game on a Graph}

\subsection{\label{subsec:prisoner}Prisoner's Dilemma}

In game theory, a strategic-form game is the standard representation of a multiplayer game which is given by the relation between the strategies and payoffs of players.
Generally, an $n$-player strategic-form game is defined by the following sets of ingredients:
\begin{align}
  \label{eq:strategic-form}
  G=(M,\set{S^i}_{i\in M}, \set{f^i}_{i\in M}),
\end{align}
where $M=\set{1,\cdots,n}$ is a set of players, $S^i$ is a set of possible strategies or actions of player $i$,
and $f^i:S^1\times\cdots\times S^n\rightarrow\mathbb{R}$ is a payoff function of player $i$.
In this game, each player $i\in\set{1,\cdots, n}$ chooses his strategy $s^i\in S^i$ without knowing the other players' choice.
Then, player $i$ is given a payoff $f^i(s^1,\cdots,s^n)$ depending on the combination of the actions of all players.
The goal of player $i$ is to maximize his own payoff.

One example is the prisoner's dilemma game \cite{william1992prisoner}.
A payoff bimatrix is generally written as in TABLE \ref{table:prisoner}.

\begin{table}[H]
\caption{Payoff matrix of the prisoner's dilemma game.}
\label{table:prisoner}
  \centering
  \begin{tabular}{c|cc}
    Alice\textbackslash Bob & C & D\\ \hline
    C & $(R,R)$ & $(S,T)$\\
    D & $(T,S)$ & $(P,P)$
  \end{tabular}
\end{table}

Here actions C and D denote cooperation and defection, either of which is chosen by each player;
$R$, $P$, $T$ and $S$ represent the reward, punishment, temptation and sucker's payoff, respectively.
In the conventional prisoner's dilemma game, the inequality $T>R>P>S$ holds.
Assume that two players, Alice and Bob, participate in the game, and they are given two choices: cooperation and defection.
Their goal is to separately choose the action that maximizes their payoff.
If both players cooperate with each other, both of them receive the reward $R$.
If both players defect, both of them receive the punishment $P$, which is lower that the reward $R$.
This means that mutual cooperation is more preferable than mutual defection for both players.
If one player unilaterally defects and the other player cooperates, the defector receives the highest payoff $T$ and the cooperator receives the lowest payoff $S$.
Since $T>R$, the players may be tempted to choose defection for a higher payoff.
Since $P>S$, the players may deviate from mutual cooperation to avoid ending up with the worst result.
Consequently, both players are likely to choose defection without any communication beforehand, although mutual cooperation would give higher payoffs.
This is why the game with the payoff bimatrix in TABLE \ref{table:prisoner} is referred to as the ``dilemma''.
From the fact that defection is Alice's best response to Bob's defection and vice versa, it is clear that this game has a Nash equilibrium (D, D).
Note that this equilibrium is not Pareto optimal.
Furthermore, defection is the dominant strategy for both players;
Defection always gives Alice a higher payoff than cooperation regardless of Bob's choice.
In this sense, we may say that it is rational for a selfish player to choose defection.
For this game to be the conventional prisoner's dilemma game in a strict sense, we have to impose the condition that $T>R>P>S$.
In this paper, however, we consider extended cases where this condition does not necessarily holds.

The donation game is a special version of the prisoner's dilemma as shown in TABLE \ref{table:donation} \cite{hilbe2013evolution}.
\begin{table}[H]
\caption{Payoff matrix of the donation game}
\label{table:donation}
  \centering
  \begin{tabular}{c|cc}
    Alice\textbackslash Bob & C & D\\ \hline
    C & $(b-c,b-c)$ & $(-c,b)$\\
    D & $(b,-c)$ & $(0,0)$
  \end{tabular}
\end{table}
\noindent
This bimatrix can be obtained simply by setting $R=b-c$, $P=0$, $T=b$ and $S=-c$ in TABLE \ref{table:prisoner}, and we have the condition that $b>c>0$ to maintain the dilemma \cite{tanimoto2007relationship}, where $b$ stands for the benefit and $c$ stands for the cost.
In this game, each player can choose two actions, cooperation and defection.
To cooperate, each player pays the cost $c$ and provides the benefit $b$ to the other.
To defect, each player does nothing to the other.
As discussed in the prisoner's dilemma game, (D, D) is the Nash equilibrium but not Pareto optimal.
The profile (C, C) is Pareto optimal when the benefit $b$ is greater than the cost $c$.

\subsection{Update Rule of Strategies}

To consider an iterated prisoner's game, we prepare a subclass of the prisoner's dilemma game.
The prisoner's dilemma game has basically four parameters, and the donation game has basically two parameters.
However, for simplicity, we here consider a one-parameter case, which is obtained by substituting $b=1+r$ and $c=-r$.

\begin{table}[H]
\caption{Payoff matrix of the one-parameter prisoner's dilemma}
\label{table:one-parameter}
  \centering
  \begin{tabular}{c|cc}
    Alice\textbackslash Bob & C & D\\ \hline
    C & $(1,1)$ & $(-r,1+r)$\\
    D & $(1+r,-r)$ & $(0,0)$
  \end{tabular}
\end{table}

In the above table, $r$ is referred to as the cost-to-benefit ratio \cite{wang2010aspiring}.
For $r>0$, this game can be regarded as the standard prisoner's dilemma, while for $-1<r<0$, mutual cooperation is promoted because the profile (C,C) is not only Pareto optimal but also the Nash equilibrium.

The setting of the game we consider is as follows.
First, construct a $k$-regular graph with a number $N$ of vertices, each of which is connected to different $k$ vertices by edges.
Then, assign players $1,\cdots,N$ to different vertices.
As an initial state, each player independently has a random strategy C or D.
We assume that each pair of two neighboring players interact with each other by iterating the game specified in TABLE \ref{table:one-parameter}.
In each step, all the players play the game with their $k$ neighbors.
Then they are given payoffs for $k$ game results according to TABLE \ref{table:one-parameter}.
The goal of each player is to maximize his averaged payoff.
After each game, one player has a chance to change his strategy according to the pairwise-Fermi update rule \cite{pacheco2006coevolution}:
one player $x\in\set{1,\cdots,N}$ is randomly chosen out of $N$ players with uniform probability.
Then player $x$ chooses one neighbor $y$ out of his $k$ neighbors with uniform probability as a player to be imitated.
Each player is assumed to know the actions and payoffs of his neighbors in the previous game, and he can utilize the knowledge to change his strategy.
Let $s_x$ be the strategy that player $x$ takes at a single step and $p_x$ be the averaged payoff that player $x$ obtains per single game.
Player $x$ can imitate the strategy of player $y$ by comparing his own payoff $p_x$ to the neighbor's payoff $p_y$ with the following probability:
\begin{align}
  \label{eq:pairwise-fermi}
  \mathrm{Pr}(s_x\leftarrow s_y) = \frac{1}{1+\exp[(p_x-p_y)/T]},
\end{align}
where $T> 0$ is referred to as the temperature.
The temperature $T$ determines the strength of natural selection.
As the temperature approaches zero, the process becomes deterministic;
if player $x$ finds that he has a payoff less than his neighbor $y$, the transition probability in Eq. (\ref{eq:pairwise-fermi}) approaches one.
If player $x$ finds that he has a payoff more than his neighbor $y$, the transition probability approaches zero.
On the other hand, as the temperature goes to infinity, the transition probability approaches one half regardless of the payoffs, and therefore the process becomes completely random.

\section{\label{sec:ensemble-averaged_square}Ensemble-Averaged Quantities of a Square Lattice}

In this section, we consider the heatmap of the ensemble-averaged quantities in the iterated prisoner's dilemma games.
Here a heatmap shows a quantity, such as the fraction of cooperators, the time-variance of the fraction and the correlation of neighboring pairs, as a function of the temperature and the cost-to-benefit ratio, where the quantity is displayed by using colors on a two-dimensional plane.
In the case where two players play a one-shot game specified by TABLE \ref{table:one-parameter} with a positive cost-to-benefit ratio $r$, they do not have an incentive to choose the cooperative action.
However, in the case where there are multiple players and they play the same game repeatedly, the emergence of mutual cooperation can be observed.
We change the cost-to-benefit ratio and the temperature to investigate the emergence of cooperation in numerical simulations.

\subsection{\label{subsec:fraction_square}Fraction of Cooperators}

\begin{figure}[htb]
  \begin{minipage}{0.99\hsize}
    \centering
    \includegraphics[width=7cm]{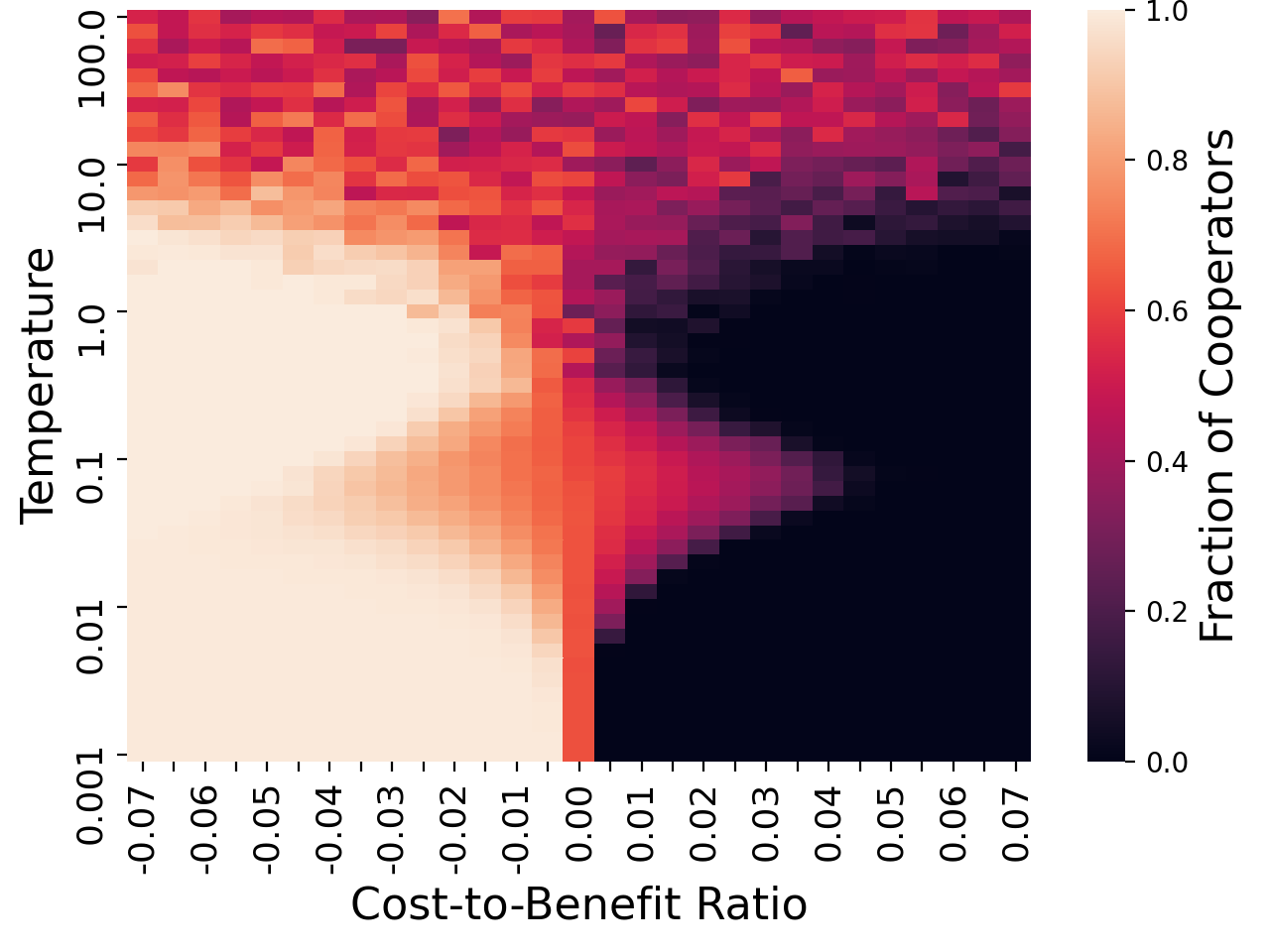}
  \end{minipage}
  \begin{minipage}{0.99\hsize}
    \centering
    \includegraphics[width=7cm]{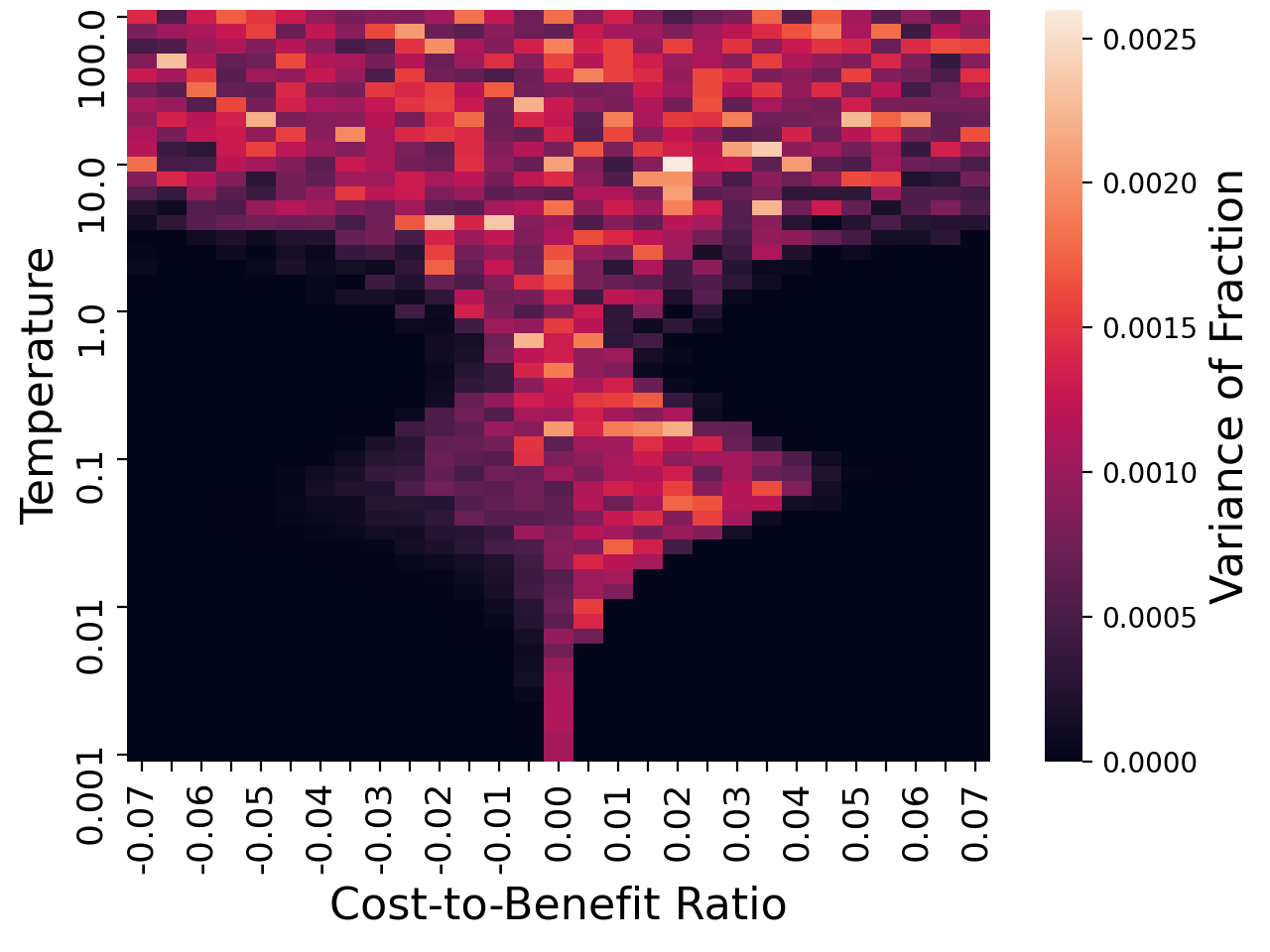}
  \end{minipage}
  \caption{
  \label{fig:heatmap_frac_square_20220614_heatmap_var_20220616}
  (Top) The fraction of cooperators appearing in the iterated prisoner's dilemma games on the square lattice.
  (Bottom) The variance of the fraction of cooperators over time appearing in the iterated prisoner's dilemma games on the square lattice.
  }
\end{figure}

The top panel in Fig. \ref{fig:heatmap_frac_square_20220614_heatmap_var_20220616} shows the fraction of cooperators that appear in the iterated game on an $L\times L$ square lattice with a size of $L=2^5$.
We have performed \(10\) independent simulations with different initial states, and for each simulation the number of time steps is \(1000\) times the number of players.
The shown result is the average over these simulations.
The neighborhood is defined as follows: when we consider a square lattice, each player is placed at a lattice point \((n, m)\), where \(n, m \in {1,...,L}\).
We also impose the periodic boundary condition on the lattice.
Then, when the distance between two players is 1, we refer to that pair of players as neighbors.
The white region corresponds to a cooperative regime in which all the players take the cooperative action.
The black region corresponds to a defective regime in which all the players take the defective action.
The red region corresponds to a mixed regime in which cooperative players and defective players coexist after a long run.
As we can see in Fig. \ref{fig:heatmap_frac_square_20220614_heatmap_var_20220616}, the mixed region consists of a diamond-shaped region and a high-temperature regime above it.

In the top panel in Fig. \ref{fig:heatmap_frac_square_20220614_heatmap_var_20220616}, as the cost-to-benefit ratio increases, the fraction of cooperators decreases.
As the cost-to-benefit ratio decreases, the fraction of cooperators increases.
We can see a boundary between the cooperative region and the mixed region in the regime of the negative cost-to-benefit ratio.
The boundary varies in a nonmonotonic way with respect to the temperature;
in the low-temperature region of $T<0.05$, the slope of the boundary is negative, while in the high-temperature region of $T>0.05$, the slope is positive.
A similar behavior can be seen in the region of the positive cost-to-benefit ratio.

We also consider the heatmap of fluctuations of the fraction of cooperators in the prisoner's dilemma game on a square lattice.
In the top panel in Fig. \ref{fig:heatmap_frac_square_20220614_heatmap_var_20220616}, we find the mutual cooperation in a certain region and the phase-transition-like behavior.
To investigate this behavior, we perform numerical simulations of the fluctuations of the fraction of cooperators over time.
The bottom panel in Fig. \ref{fig:heatmap_frac_square_20220614_heatmap_var_20220616} shows the result of the numerical simulation.
Here we define the variance of the fraction of cooperators over time as
\begin{align}
  \label{eq:variance-regular}
  \frac{1}{\tau}\sum_{t=1}^\tau
  \left[
    \frac{1}{N}\sum_{i=1}^N x^i_t
    - \frac{1}{\tau}\sum_{t^{\prime}=1}^\tau \frac{1}{N}\sum_{i=1}^N x^i_{t^{\prime}}
  \right]^2,
\end{align}
where $i\in\set{1,\cdots,N}$ denotes each player, $\braket{ij}$ denotes each pairs of neighboring players,
$t$ denotes each step when numerical results are obtained to calculate the variance of the fraction,
and \(\tau\) denotes the number of time steps in the whole iterated game.

It is interesting that the heatmap shown in Fig. \ref{fig:heatmap_frac_square_20220614_heatmap_var_20220616} is analogous to a phase diagram concerning cooperation and defection.

\subsection{\label{subsec:correlation_square}Correlation of Neighboring Pairs}

We also consider the heatmap of the correlation of neighboring pairs in the prisoner's dilemma game on a square lattice.
Here by the correlation we mean the correlation between the strategies of two players placed at vertices connected by edges on the graph.
The correlation $C$ is defined as follows:

\begin{align}
  \label{eq:correlation_square}
  C:=
  \frac{1}{2N}\sum_{(ij)}
  \left[
    \braket{x_ix_j}-\braket{x_i}\braket{x_j}
  \right],
\end{align}
where $x_i\in\set{0,1}$ represents the strategy of player $i\in\set{1,\cdots,N}$;
$x_i=0$ if player $i$ is a cooperator and $x_i=1$ if player $i$ is a defector.
The sum is taken over all neighboring pairs \((ij)\), which is divided by the number $2N$ of neighboring pairs, and
$\braket{\cdot}$ represents the time average.

\begin{figure}[htb]
  \centering
  \includegraphics[width=8cm]{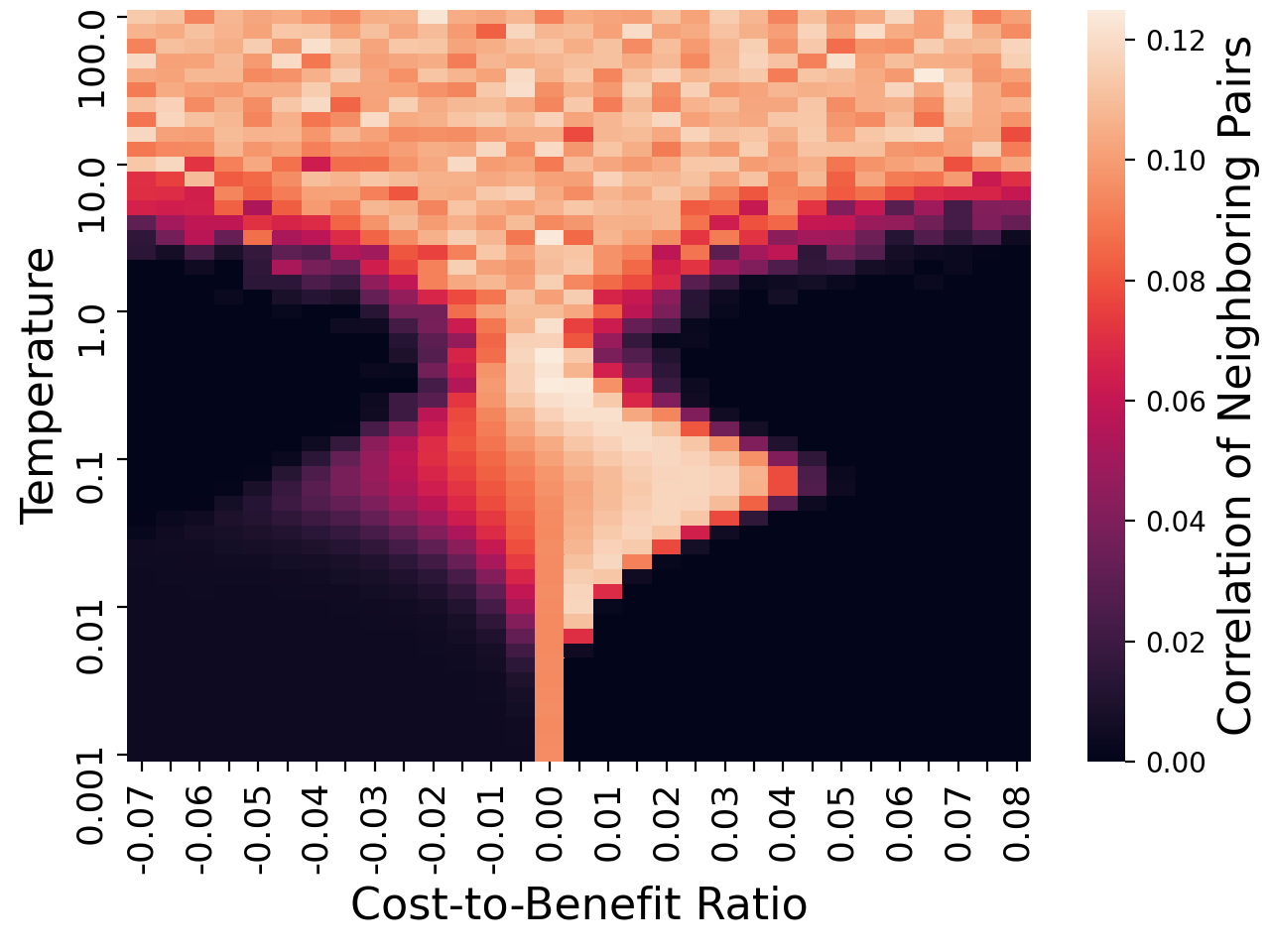}
  \caption{
  \label{fig:heatmap_cor_square_20220616}
  Correlation $C$ of the strategies of neighboring pairs in the iterated prisoner's dilemma games on the square lattice,
  where $C$ is defined by Eq. (\ref{eq:correlation_square}).
  }
\end{figure}

We have seen the phase-transition-like behavior and the fluctuations of the fraction of cooperators in the previous subsection.
We now investigate the relation between the correlation of players and the fraction of cooperators.
In Fig. \ref{fig:heatmap_cor_square_20220616}, we find strong correlation between neighboring pairs in the mixed region.
We observe a diamond-shaped region where the correlation is significantly enhanced.
This region belongs to the mixed region discussed earlier.

It is expected that in the mixed region with higher temperature $T\gtrsim 1.0$ above the diamond region,
the correlation takes on a small value after being averaged over a sufficiently large number of samples.
This is because this is a high-temperature region, and thus the fluctuation in this region should be larger than that in the low-temperature diamond region.
In the diamond region, we observe that the correlation tends to be stronger in the region with a positive cost-to-benefit ratio than in the region with a negative cost-to-benefit ratio.
The reason for this is as follows.
From TABLE \ref{table:one-parameter}, when the cost-to-benefit ratio $r$ is positive, one's payoff is maximized to be $1+r$ for unilateral defection and minimized to be $-r$ for unilateral cooperation.
Thus, every player is tempted to choose defection, and one wants to avoid being unilaterally defected by the other player to obtain a higher payoff.
Hence one needs to carefully observe the strategies of the other players and take his own strategy accordingly.
As a result, the correlation becomes stronger.
On the other hand, for the cost-to-benefit ratio $-1<r<0$, one's payoff is maximized to be $1$ for mutual cooperation and minimized to be $0$ for mutual defection.
This means that every player can maximize his own payoff simply by choosing cooperation.
Hence it is not necessary for players to carefully observe which strategy the other player takes, and therefore the correlation becomes weaker than in the case of $r>0$.

\section{\label{sec:ensemble-averaged_regular}Ensemble-Averaged Quantities of a regular graph}

In this section, we consider the heatmap of cooperation and defection in the iterated prisoner's dilemma game on a regular graph.
As shown below, mutual cooperation emerges in a certain region similarly to the case of the game on a square lattice.
We change two parameters, the cost-to-benefit ratio $r$ and the temperature $T$, to investigate a phase-transition-like behavior by numerical simulations.
Compared with the case of a square lattice, we can vary the degree $k$ of the graph, where $k$ is the number of edges that connect a vertex to other vertices.

\subsection{\label{subsec:fraction_regular}Fraction of Cooperators}

\begin{figure}[H]
  \begin{minipage}{0.49\linewidth}
    \centering
    \includegraphics[keepaspectratio, width=1.0\linewidth]{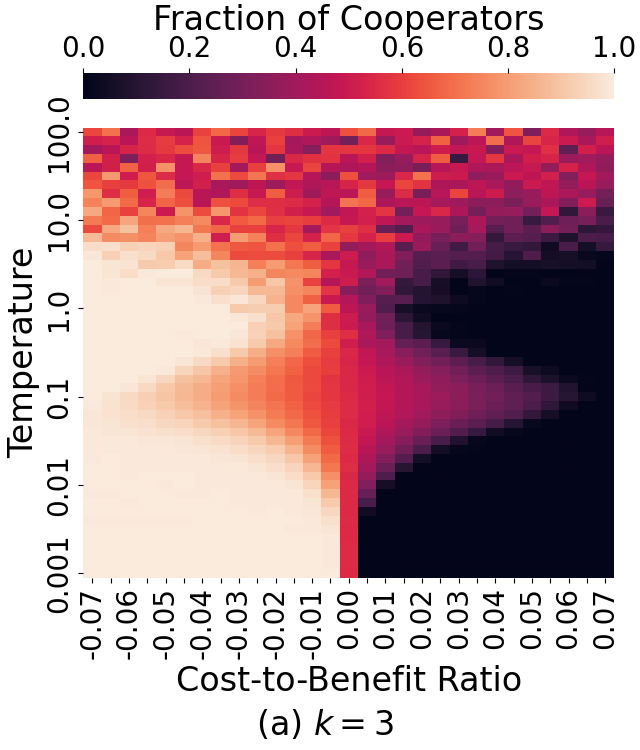}
  \end{minipage}
  \begin{minipage}{0.49\linewidth}
    \centering
    \includegraphics[keepaspectratio, width=1.0\linewidth]{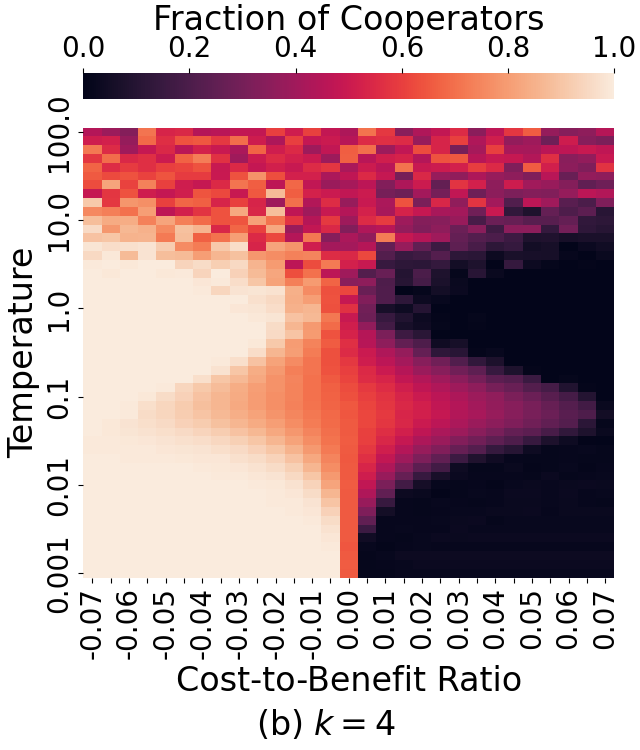}
  \end{minipage}
  \begin{minipage}{0.49\linewidth}
    \centering
    \includegraphics[keepaspectratio, width=1.0\linewidth]{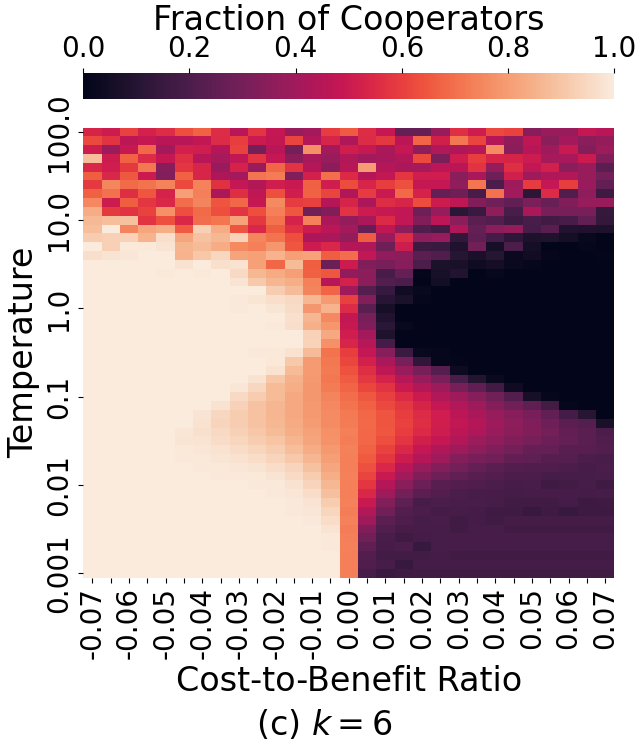}
  \end{minipage}
  \begin{minipage}{0.49\linewidth}
    \centering
    \includegraphics[keepaspectratio, width=1.0\linewidth]{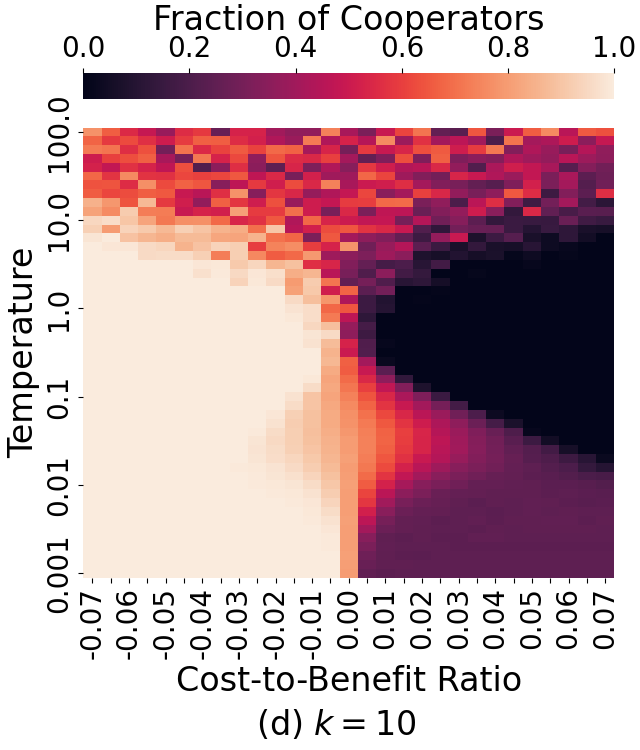}
  \end{minipage}
  \begin{minipage}{0.49\linewidth}
    \centering
    \includegraphics[keepaspectratio, width=1.0\linewidth]{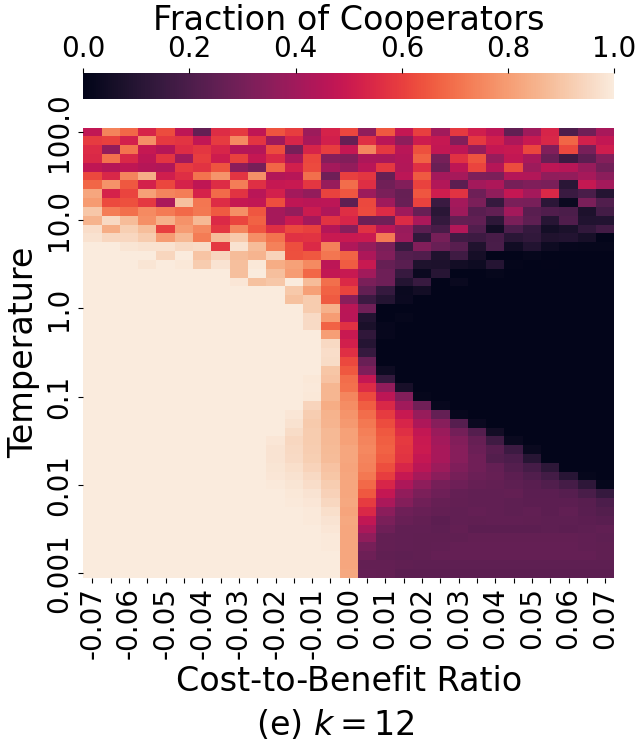}
  \end{minipage}
  \begin{minipage}{0.49\linewidth}
    \centering
    \includegraphics[keepaspectratio, width=1.0\linewidth]{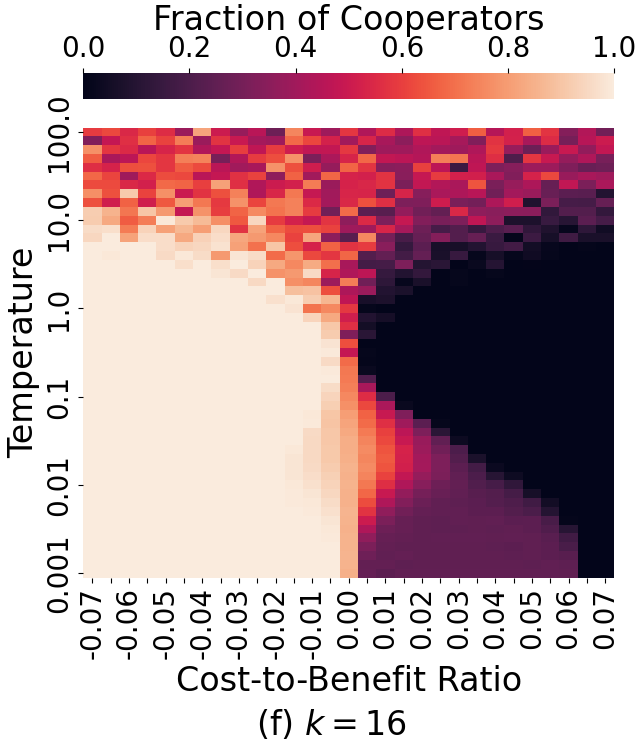}
  \end{minipage}
  \begin{minipage}{0.49\linewidth}
    \centering
    \includegraphics[keepaspectratio, width=1.0\linewidth]{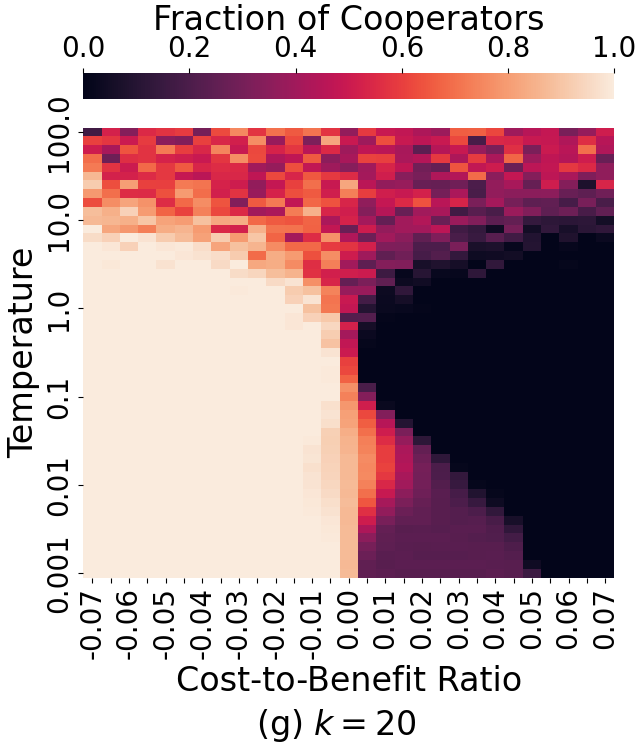}
  \end{minipage}
  \begin{minipage}{0.49\linewidth}
    \centering
    \includegraphics[keepaspectratio, width=1.0\linewidth]{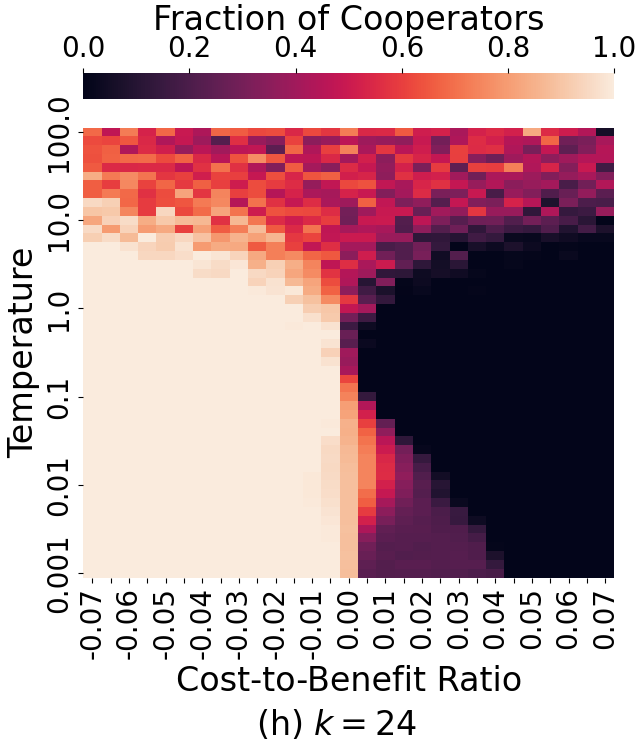}
  \end{minipage}
  \caption{
  \label{fig:heatmap_frac_regular}
  The fraction of cooperators appearing in the iterated prisoner's dilemma game on regular graphs.
  The color of each pixel represents the fraction of cooperators.
  On a $k$-regular graph, each vertex has equal edges that connect each vertex to other $k$ vertices.
  Each figure corresponds to the case of $k=3, 4, 6, 10, 12, 16, 20, 24$ from the top left panel to the bottom right panel.
  In all the graphs the number of vertices is $2^{10}$.
  A single player is placed at each vertex on the graph.
  Each player changes his own strategy over time by using the pairwise-Fermi update rule.
  }
\end{figure}

Figure \ref{fig:heatmap_frac_regular} shows the result of the numerical simulation of the iterated prisoner's dilemma game on regular graphs.
The regular graphs are randomly generated by specifying the number $N$ of vertices and the degree $k$.
We set the number of vertices to be $N=2^{10}$ and the degree to be $k=3, 4, 6, 10, 12, 16, 20, 24$.
A single player is assigned to each vertex on the graph.
Each player changes his own strategy over time by using the pairwise-Fermi update rule.
Each panel in Fig. \ref{fig:heatmap_frac_regular} shows the fraction of cooperators that appear in the iterated game on the regular graph.

We find from Fig. \ref{fig:heatmap_frac_regular} that the game result depends on the cost-to-benefit ratio $r$ and the temperature $T$.
As in the case of a square lattice discussed in the previous section, the white region shows a cooperative regime in which all the players take the cooperative action.
On the other hand, the black region shows a defective regime in which all the players take the defective action.
The red region shows a mixed regime in which cooperative players and defective players coexist after a long time.

For each given temperature $T$, as the cost-to-benefit ratio increases, the fraction of cooperators decreases.
As the cost-to-benefit ratio decreases, the fraction of cooperators increases.
We see a boundary between the cooperative region and the mixed region for the negative cost-to-benefit ratio.
The boundary varies in a nonmonotonic manner against the temperature.
In the top right panel in Fig. \ref{fig:heatmap_frac_regular}, which corresponds to the degree of $k=4$, in the low-temperature region of $T\lesssim 0.05$, the slope of the boundary is positive for the positive cost-to-benefit ratio $r>0$.
In the intermediate-temperature region of $0.05\lesssim T\lesssim 1.0$, the slope is negative.
In the high-temperature region of $T\gtrsim 1.0$, the slope is positive.
A similar nonmonotonic behavior can be seen in the region of the negative cost-to-benefit ratio.

So far the results are essentially the same as the case of a square lattice.
In Fig. \ref{fig:heatmap_frac_regular}, we observe that as the degree $k$ increases from $k=3$ to $k=6$, the boundaries tend to shift to the positive direction of the cost-to-benefit ratio.
This tendency suggests the following.
In general, the payoff matrix tells us that the mutual cooperation provides a payoff more than mutual defection.
Therefore one's cooperation tends to induce the other player's cooperation.
If the degree $k$ is large, one's temptation to choose the defective action for a higher payoff is expected to decrease.
However, as the degree increases from $k=10$ to $k=24$, the behavior of the boundaries changes;
the diamond-shaped region shrinks along the horizontal axis toward the vertical axis of zero cost-to-benefit ratio $r=0$, and the slope of boundaries surrounding the mixed region become steeper.
Since the diamond-shaped region is the region in which the fraction of cooperators fluctuates,
this result may be interpreted as the larger mean-field effect for larger $k$.
In other words, as the degree \(k\) increases, the network structure becomes closer to the case of a well-mixed population,
where natural selection prevents the coexistence of cooperators and defectors.

Another finding concerns the regime located in the lower-right corner of the diamond-shaped mixed region in the panels in Fig. \ref{fig:heatmap_frac_regular};
for a small degree $k$, this region exhibits nearly extinct cooperators,
while for a large degree $k$, this region involves a relatively large number of cooperators in comparison with the case of a small degree $k$.
In contrast, in the upper-right corner of the diamond-shaped region in the panels in Fig. \ref{fig:heatmap_frac_regular}, this tendency is not seen;
the regime with intermediate temperature $T$ and positive cost-to-benefit ratio $r$ is almost unchanged regarding to the degree $k$ of the regular graph.
Such temperature-dependent behavior reminds us of the difference between gradient descent (GD) and stochastic gradient descent (SGD);
in GD, the optimization process is deterministic and likely to be trapped in a local optimal point,
while in SGD, the optimization process is probabilistic and likely to reach the global optimal point.
This difference is analogous to that of a stochastic process which obeys the pairwise-Fermi update rule in Eq. (\ref{eq:pairwise-fermi}) for the lower-temperature case and the process for the higher-temperature case;
the lower-temperature process is deterministic and tends to eliminate cooperators for positive cost-to-benefit ratio \(r > 0\),
while the higher-temperature process is probabilistic and tends to maintain the coexistence of cooperators and defectors.

Additionally, we observe a quantitative difference between a square lattice and a 4-regular graph according to the heatmap of the fraction of cooperators;
comparing the top panel of Fig. \ref{fig:heatmap_frac_square_20220614_heatmap_var_20220616} and Fig. \ref{fig:heatmap_frac_regular} (b),
the width of the diamond-shaped region along the horizontal axis for a 4-regular graph is wider than that for a square lattice.
A possible reason for this observation is as follows.
When players are placed on a lattice, it is more likely that clusters of cooperators are formed and maintained for a long time.
In contrast, when players are placed at a regular graph that is randomly generated, we do not expect clusters of cooperators to be formed.

\subsection{\label{subsec:correlation_regular}Correlation of Neighboring Pairs}

We also consider the heatmap of the correlation of neighboring pairs in the prisoner's dilemma game on $k$-regular graphs.
The correlation $C$ is defined as
\begin{align}
  \label{eq:correlation_regular}
  C:=
  \frac{2}{kN}\sum_{(ij)}
  \left[
    \braket{x_ix_j}-\braket{x_i}\braket{x_j}
  \right],
\end{align}
where $x_i\in\set{0,1}$ represents the strategy of player $i\in\set{1,\cdots,N}$;
$x_i=0$ if player $i$ is a cooperator and $x_i=1$ if player $i$ is a defector.
The sum is taken over all neighboring pairs \((ij)\), $kN/2$ is the number of neighboring pairs, and
$\braket{\cdot}$ represents the time average.
Here we refer to neighboring pairs as players located at vertices that are connected by edges on a graph.

\begin{figure}[H]
  \begin{minipage}{0.49\linewidth}
    \centering
    \includegraphics[keepaspectratio, scale=0.25]{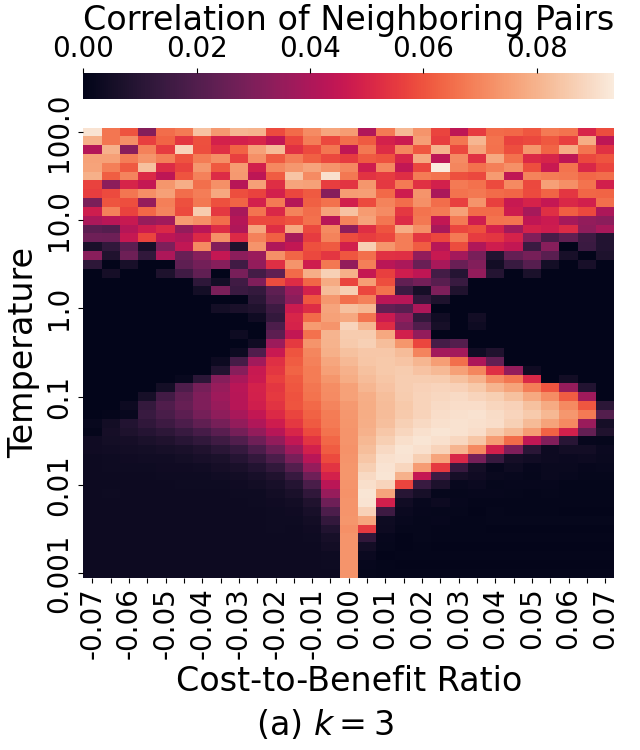}
  \end{minipage}
  \begin{minipage}{0.49\linewidth}
    \centering
    \includegraphics[keepaspectratio, scale=0.25]{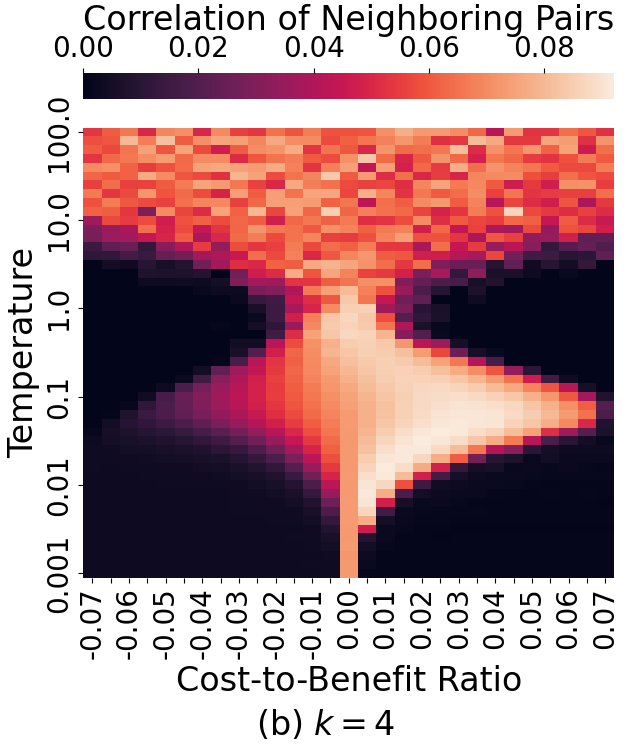}
  \end{minipage}
  \begin{minipage}{0.49\linewidth}
    \centering
    \includegraphics[keepaspectratio, scale=0.25]{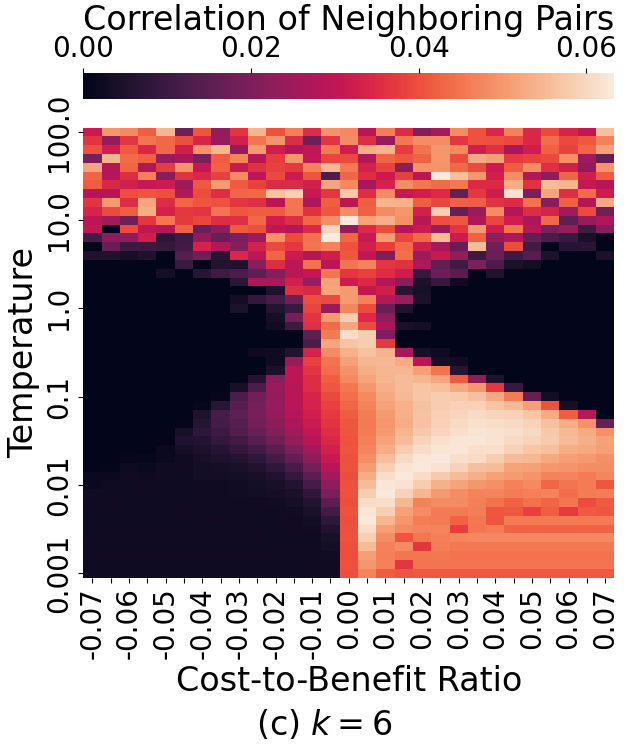}
  \end{minipage}
  \begin{minipage}{0.49\linewidth}
    \centering
    \includegraphics[keepaspectratio, scale=0.25]{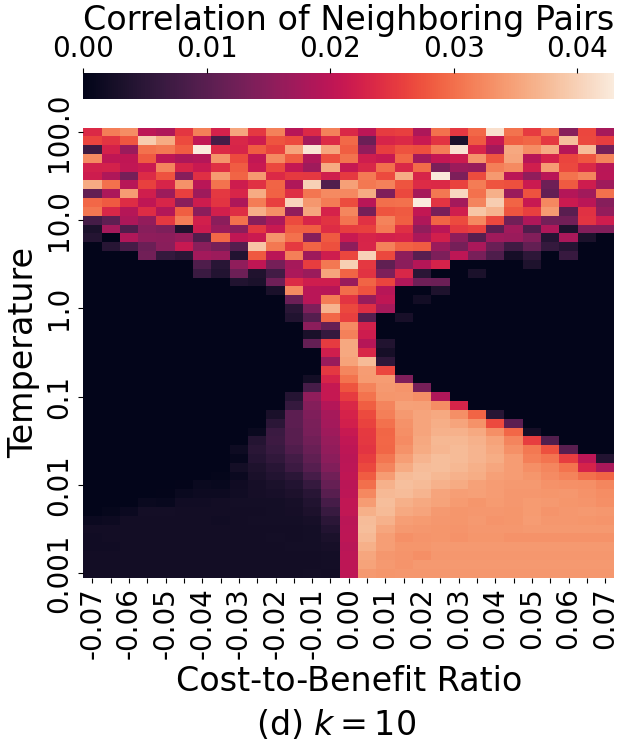}
  \end{minipage}
  \begin{minipage}{0.49\linewidth}
    \centering
    \includegraphics[keepaspectratio, scale=0.25]{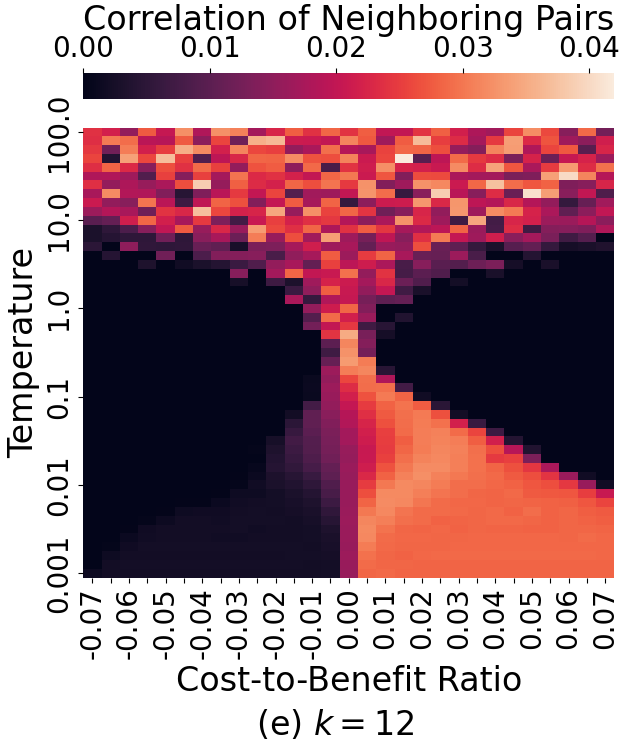}
  \end{minipage}
  \begin{minipage}{0.49\linewidth}
    \centering
    \includegraphics[keepaspectratio, scale=0.25]{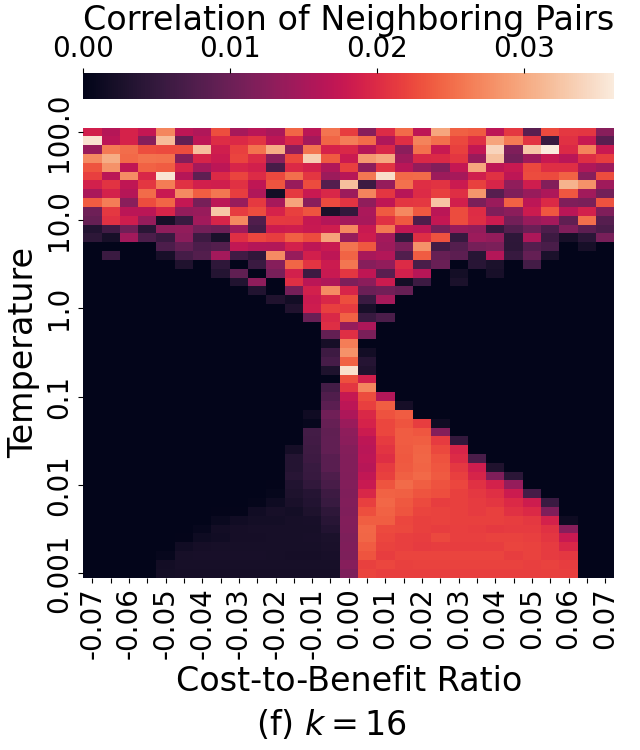}
  \end{minipage}
  \begin{minipage}{0.49\linewidth}
    \centering
    \includegraphics[keepaspectratio, scale=0.25]{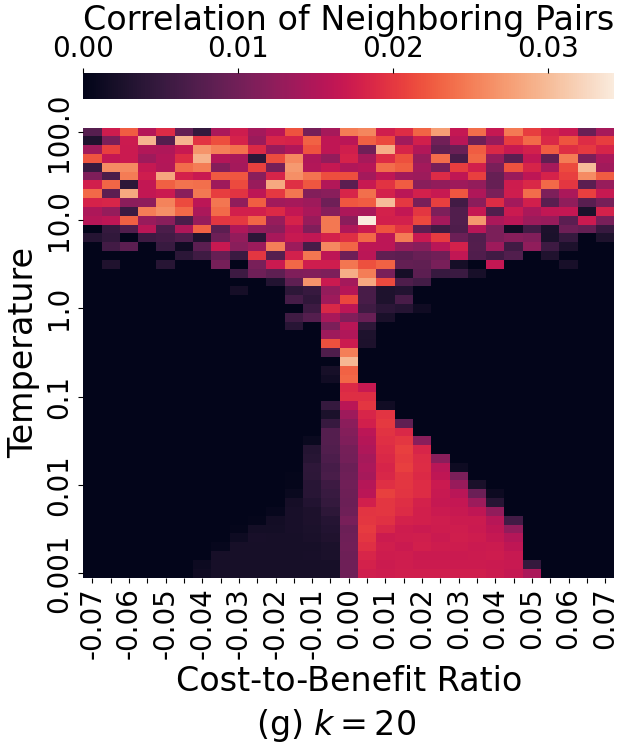}
  \end{minipage}
  \begin{minipage}{0.49\linewidth}
    \centering
    \includegraphics[keepaspectratio, scale=0.25]{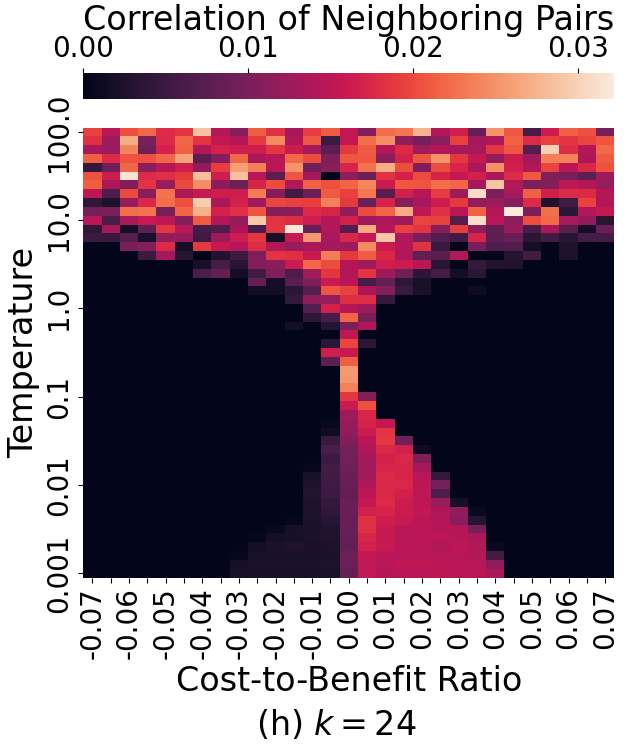}
  \end{minipage}
  \caption{
  \label{fig:heatmap_cor_regular}
  Correlation $C$ of the strategies of neighboring pairs in the iterated prisoner's dilemma game on regular graphs.
  The color of each pixel represents the fraction of cooperators.
  On a $k$-regular graph, each vertex has equal edges that connect each vertex to other $k$ vertices.
  Each figure shows the case of $k=3, 4, 6, 10, 12, 16, 20, 24$ from the top left panel to the bottom right one.
  In all the graphs the number of vertices is $2^{10}$.
  A single player is placed at each vertex on the graph.
  Each player changes his own strategy over time according to the pairwise-Fermi update rule.
  }
\end{figure}

Figure \ref{fig:heatmap_cor_regular} shows the correlation of strategies of neighboring pairs in the iterated prisoner's dilemma game on a regular graphs.
Each panel corresponds to the numerical result on a random regular graph which has the number $N=2^{10}$ of vertices and the degree $k = 3, 4, 6, 10, 12, 16, 20, 24$.
The landscape of the heatmap of the correlation in Fig. \ref{fig:heatmap_cor_regular} resembles that of the fraction of cooperators in Fig. \ref{fig:heatmap_frac_regular}, similarly to the case of a square lattice.
Noting that the color bar of a heatmap in Fig. \ref{fig:heatmap_cor_regular} depends on the degree of the graph,
we find that as the degree increases, the correlation of neighboring pairs decreases.
This is because a player surrounded by an increasing number of neighbors finds the per capita influence of his neighbors to be averaged out.

\section{\label{sec:single}Single-Trajectory Quantities}

In this section, we consider the detailed dynamics of the iterated prisoner's dilemma game.
Specifically, we track individual real-time trajectories of the fraction of cooperators in an iterated game.
Then we analyze the power spectrum of the trajectories, which turns out to be approximated as a superposition of several sine waves with low frequencies.
In particular, we find that the spectral density $S(f)$ generally follows the power-law distribution of the pink noise in such a manner that $S(f)\propto f^{-\alpha}$, where $f$ denotes the frequency.
Furthermore, the exponent $\alpha$ is found to be close to one.

\subsection{\label{subsec:realtime}Real-Time Dynamics of Fraction of Cooperators}

\begin{figure}[htb]
  \begin{minipage}{0.99\linewidth}
    \centering
    \includegraphics[keepaspectratio, scale=0.25]{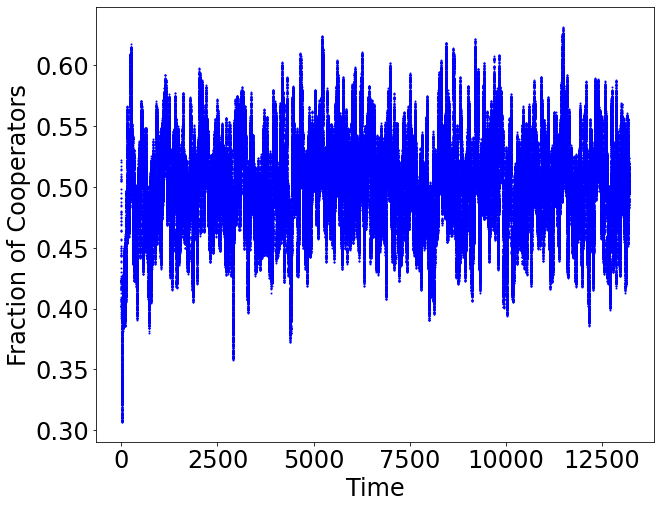}
  \end{minipage}
  \begin{minipage}{0.99\linewidth}
    \centering
    \includegraphics[keepaspectratio, scale=0.25]{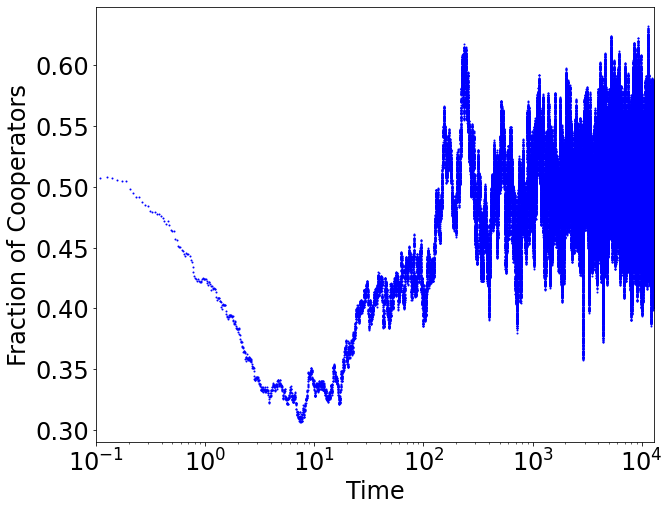}
  \end{minipage}
  \caption{
  \label{fig:2022-09-20-19-00-00_L32_K0.1_r0.02_regular_deg4_seed42_linear_log}
  Real-time trajectory of the fraction of cooperators in an iterated prisoner's dilemma game on a $4$-regular graph with cost-to-benefit ratio $r=0.02$ and temperature $T=0.1$.
  This pair of parameters belongs to the mixed region in Fig. \ref{fig:heatmap_frac_regular}.
  Note that the time scale of the bottom panel is shown on a logarithmic scale.
  }
\end{figure}

Figure \ref{fig:2022-09-20-19-00-00_L32_K0.1_r0.02_regular_deg4_seed42_linear_log} shows the real-time trajectory of the fraction of cooperators in an iterated prisoner's dilemma game on a $4$-regular graph, where the cost-to-benefit ratio is set to be $r=0.02$, the temperature is set to be $T=0.1$ and the number of players is set to be $N=2^{10}$.
Throughout this paper, time is measured in units of \(1/N\).
Here, the pairwise-Fermi update rule is applied every time $2n$ prisoner's dilemma games are performed.
In Fig. \ref{fig:2022-09-20-19-00-00_L32_K0.1_r0.02_regular_deg4_seed42_linear_log}, the unit time corresponds to $N$ pairwise-Fermi updates.
In the top panel, we see that the curve oscillates around the fraction of one half.
This implies that the set of parameters $r=0.02$ and $T=0.1$ belongs to the mixed region.
In the bottom panel, however, we see that the fraction of cooperators decreases in the early stage.
This is thought to be due to the following reason.
In the initial state, each player is independently given an action, i.e., cooperation or defection.
Cooperators are exploited by their neighboring defectors because of the positive cost-to-benefit ratio, and thus some cooperators change their strategy to defection by following the update rule.
Meanwhile, the fraction of cooperators recovers after $10$ time units.

\begin{figure}[htb]
  \begin{minipage}{0.99\linewidth}
    \centering
    \includegraphics[keepaspectratio, scale=0.25]{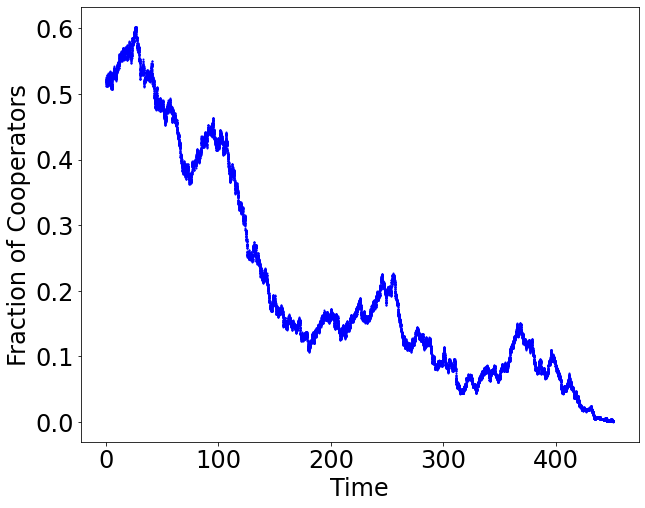}
  \end{minipage}
  \caption{
  \label{fig:2022-09-20-19-00-00_L32_K1.0_r0.02_regular_deg4_seed42_linear}
  Real-time trajectory of the fraction of cooperators in an iterated prisoner's dilemma game on a $4$-regular graph with cost-to-benefit ratio $r=0.02$ and temperature $T=1.0$.
  The fraction of cooperators eventually vanishes at about $450$ time units, since the pair of parameters belongs to the defective region in Fig. \ref{fig:heatmap_frac_regular}.
  }
\end{figure}

Figure \ref{fig:2022-09-20-19-00-00_L32_K1.0_r0.02_regular_deg4_seed42_linear} shows the real-time trajectory of the fraction of cooperators in an iterated prisoner's dilemma game on a $4$-regular graph, where the cost-to-benefit ratio is $r=0.02$, the temperature is $T=1.0$ and the number of players is $N=2^{10}$.
The fraction of cooperators is initially about one half before the updates similarly to the case of $T=0.1$ in Fig. \ref{fig:2022-09-20-19-00-00_L32_K0.1_r0.02_regular_deg4_seed42_linear_log}.
In this case, however, cooperators go extinct during the simulation, and the strategies of players are no longer updated after the fraction becomes zero at about $450$ time units.
Moreover, this means that the set of parameters $r=0.02$ and $T=1.0$ belongs to the defective region.

\begin{figure}[htb]
  \begin{minipage}{0.99\linewidth}
    \centering
    \includegraphics[keepaspectratio, scale=0.25]{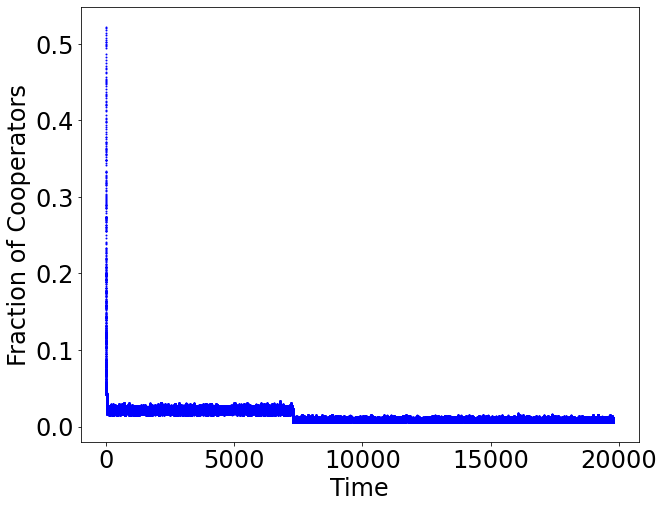}
  \end{minipage}
  \begin{minipage}{0.99\linewidth}
    \centering
    \includegraphics[keepaspectratio, scale=0.25]{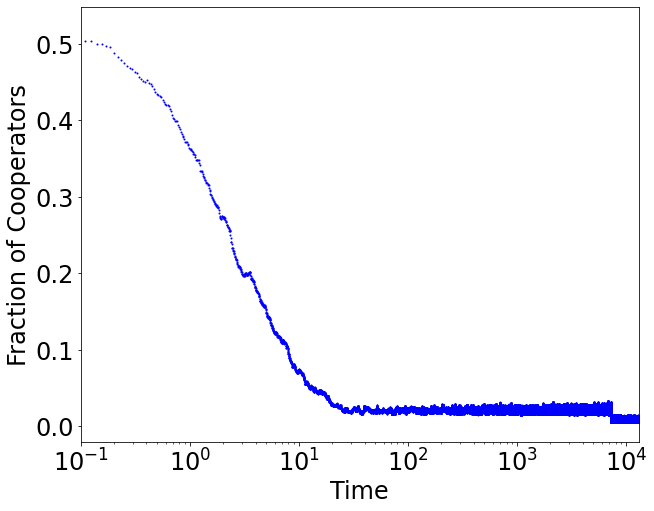}
  \end{minipage}
  \caption{
  \label{fig:2022-09-20-19-00-00_L32_K0.0032_r0.02_regular_deg4_seed42_linear_log}
  Real-time trajectory of the fraction of cooperators in an iterated prisoner's dilemma game on a $4$-regular graph with cost-to-benefit ratio $r=0.02$ and temperature $T=3.2\times 10^{-3}$.
  Note that the time scale of the bottom panel is shown on a logarithmic scale.
  }
\end{figure}

Figure \ref{fig:2022-09-20-19-00-00_L32_K0.0032_r0.02_regular_deg4_seed42_linear_log} shows the real-time trajectory of the fraction of cooperators in an iterated prisoner's dilemma game on a $4$-regular graph with cost-to-benefit ratio $r=0.02$
and temperature is $T=3.2\times 10^{-3}$.
We see that the fraction of cooperators is initially about one half as in Fig. \ref{fig:2022-09-20-19-00-00_L32_K0.1_r0.02_regular_deg4_seed42_linear_log}
and Fig. \ref{fig:2022-09-20-19-00-00_L32_K1.0_r0.02_regular_deg4_seed42_linear}.
However, the fraction rapidly decays in the initial stage, and thereafter the fraction remains to be small.
The reason for the rapid decay is as follows.
At low temperature, the process following the pairwise-Fermi update rule becomes almost deterministic;
the probability that an advantageous strategy is chosen approaches one, while the probability that a disadvantageous strategy is chosen approaches zero for each update.
For a positive cost-to-benefit ratio, therefore, defection becomes dominant almost without fluctuations in the early phase.

From Fig. \ref{fig:2022-09-20-19-00-00_L32_K0.0032_r0.02_regular_deg4_seed42_linear_log}, we find that at extremely low temperature the fraction of cooperators quickly decreases in $10$ time units.
However, it is notable that the fraction stays non-zero values even after a long time.
This means that the set of parameters $r=0.02$ and $T=3.2\times 10^{-3}$, which is located in the lower-right area below the diamond-shaped region in the upper-right panel in Fig. \ref{fig:heatmap_frac_regular}, belongs to the mixed region.
Furthermore, this result is consistent with the fact that the fraction is clearly nonzero in the lower-right area with degree $k\geq 6$ in Fig. \ref{fig:heatmap_frac_regular}.

\begin{figure}[htb]
  \begin{minipage}{0.99\linewidth}
    \centering
    \includegraphics[keepaspectratio, scale=0.25]{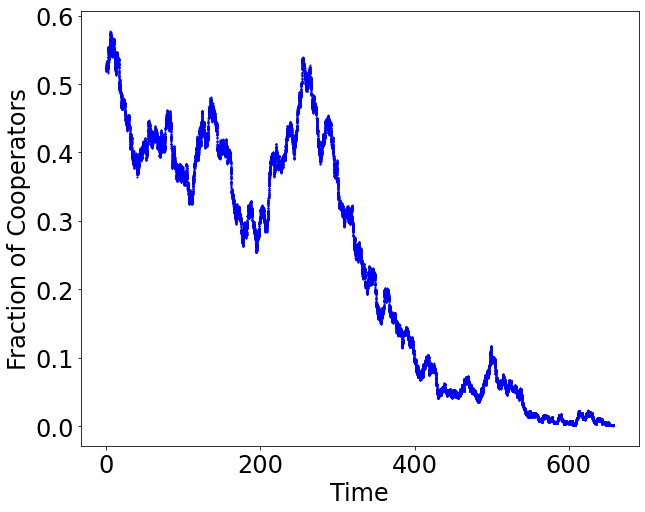}
  \end{minipage}
  \caption{
  \label{fig:2022-09-20-20-38-18_L32_K100.0_r0.02_regular_deg4_seed42}
  Real-time trajectory of the fraction of cooperators in an iterated prisoner's dilemma game on a $4$-regular graph with cost-to-benefit ratio $r=0.02$ and temperature $T=100.0$.
  Note that when we perform multiple independent numerical simulations with different random seeds,
  some numerical results show that cooperators go extinct, as shown in this figure,
  while others show that defectors go extinct (not shown).
  }
\end{figure}

\subsection{\label{subsec:spectrum}Power Spectrum: Pink Noise}

We discuss the power spectrum of real-time trajectories of the fraction of cooperators in an iterated prisoner's dilemma game on a regular graph.
The fraction of cooperators is recorded at every time interval $\delta t$, and thus the fraction at each time can be labeled by assigning an index $m = 0,\cdots,N-1$;
the fraction of cooperators at time $m\delta t$ is denoted by $a_m$ $(m=0,\cdots,N-1)$,
where $N$ is the number of the points in the data set.
Let the discrete Fourier transform $A_k$ $(k=0,\cdots,N-1)$ of $a_m$ be defined as
\begin{align}
  \label{eq:Fourier-transform}
  A_k = \sum_{m=0}^{N-1} a_m \exp\left(-i\frac{2\pi mk}{N}\right),
\end{align}
and the frequency $f_k$ be defined as
\begin{align}
  \label{eq:frequency_fk}
  f_k = k/(N\delta t).
\end{align}
Then, the original data $a_m$ can be expressed as
\begin{align}
  a_m = \frac{1}{N}\sum_{k=0}^{N-1} A_k \exp\left(i\frac{2\pi mk}{N}\right).
\end{align}
In the following, we examine the power spectrum $\abs{A_k}^2$ against frequency $f_k$ for $k \leq N/2$.

\begin{figure}[htb]
  \begin{minipage}{0.99\linewidth}
    \centering
    \includegraphics[keepaspectratio, scale=0.25]{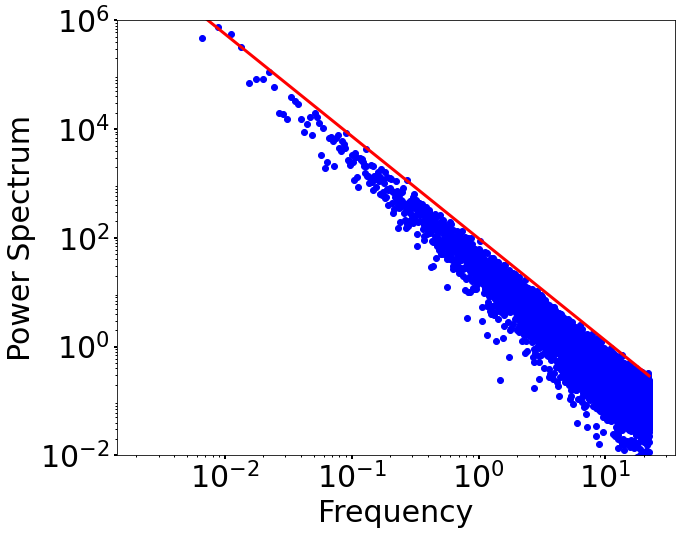}
  \end{minipage}
  \caption{
  \label{fig:spectrum_2022-09-17-22-14-13_L32_K1.0_r0.02_regular_deg4_seed42}
  Power spectrum of the real-time trajectory of the fraction of cooperators in an iterated prisoner's dilemma game on a $4$-regular graph with cost-to-benefit ratio $r=0.02$ and temperature $T=1.0$.
  The solid line is a power-law curve $S(f)\propto 1/f^{\alpha}$ with an exponent $\alpha=1.88$.
  }
\end{figure}

Figure \ref{fig:spectrum_2022-09-17-22-14-13_L32_K1.0_r0.02_regular_deg4_seed42} shows the power spectrum of the real-time trajectory
in Fig. \ref{fig:2022-09-20-19-00-00_L32_K1.0_r0.02_regular_deg4_seed42_linear} with cost-to-benefit ratio $r=0.02$ and temperature $T=1.0$.
The blue dots represent the power spectrum calculated from Eq. (\ref{eq:Fourier-transform}),
and the red solid line is a power-law curve which decays with respect to the frequency $f$ as $S(f)\propto 1/f^{\alpha}$ where the exponent is $\alpha=1.88$.
The upper envelope of the sets of the blue dots can be fitted by the red solid line.
From this result, we see that the real-time trajectory in Fig. \ref{fig:2022-09-20-19-00-00_L32_K1.0_r0.02_regular_deg4_seed42_linear} behaves like a pink noise.
Moreover, in Fig. \ref{fig:spectrum_2022-09-17-22-14-13_L32_K1.0_r0.02_regular_deg4_seed42}, a few dots within $f<0.02$ are dominant in the power spectrum.
This means that the real-time trajectory can roughly be reproduced by a superposition of a small number of low-frequency waves.

\begin{figure}[htb]
  \begin{minipage}{0.99\linewidth}
    \centering
    \includegraphics[keepaspectratio, scale=0.25]{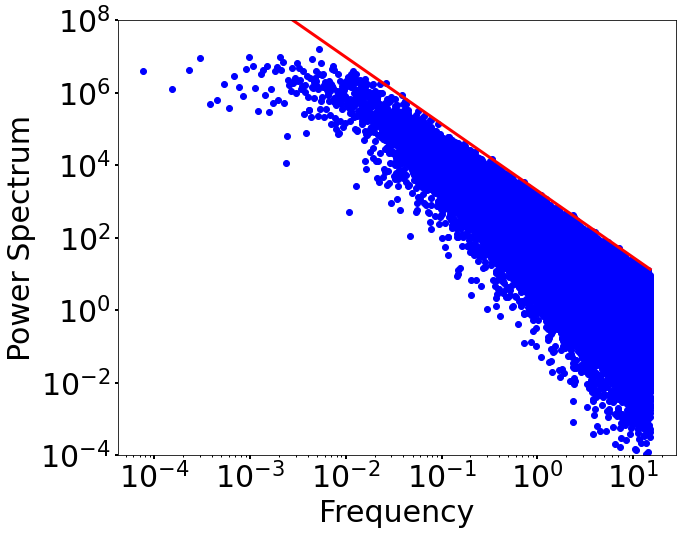}
  \end{minipage}
  \caption{
  \label{fig:spectrum_2022-09-20-19-00-00_L32_K0.1_r0.02_regular_deg4_seed42}
  Power spectrum of the real-time trajectory of the fraction of cooperators in an iterated prisoner's dilemma game on a $4$-regular graph with cost-to-benefit ratio $r=0.02$ and temperature $T=0.1$.
  The solid line is a power-law curve $S(f)\propto 1/f^{\alpha}$ with an exponent $\alpha=1.84$.
  }
\end{figure}

Figure \ref{fig:spectrum_2022-09-20-19-00-00_L32_K0.1_r0.02_regular_deg4_seed42} shows the power spectrum of the real-time trajectory
in Fig. \ref{fig:2022-09-20-19-00-00_L32_K0.1_r0.02_regular_deg4_seed42_linear_log} with cost-to-benefit ratio $r=0.02$ and temperature $T=0.1$.
As in Fig. \ref{fig:spectrum_2022-09-17-22-14-13_L32_K1.0_r0.02_regular_deg4_seed42}, the blue dots represent power spectrum, and the red solid line is proportional to $S(f)\propto 1/f^{\alpha}$ with the exponent $\alpha=1.84$.
We note that the data dots distribute over a wider range of frequency in Fig. \ref{fig:spectrum_2022-09-20-19-00-00_L32_K0.1_r0.02_regular_deg4_seed42}
than that in Fig. \ref{fig:spectrum_2022-09-17-22-14-13_L32_K1.0_r0.02_regular_deg4_seed42}.
This is because the original data obtained for $T=0.1$ have a larger number of data points than that for $T=1.0$ due to the decay of the fraction of cooperators for $T=1.0$,
and thus the available data points extend to a lower-frequency range, following Eq. (\ref{eq:frequency_fk}).
In the high-frequency region $f\gtrsim 10^{-2}$, the power spectrum decreases so that the uppermost data points distribute along the red solid line, similarly to Fig. \ref{fig:spectrum_2022-09-17-22-14-13_L32_K1.0_r0.02_regular_deg4_seed42}.
In the low-frequency region $f\lesssim 10^{-2}$, in contrast, the blue dots deviate from the red solid line.
A possible reason for this deviation from the red line is the strong correlation of neighboring pairs discussed in Sec. \ref{subsec:correlation_regular}.

\begin{figure}[htb]
  \begin{minipage}{0.99\linewidth}
    \centering
    \includegraphics[keepaspectratio, scale=0.25]{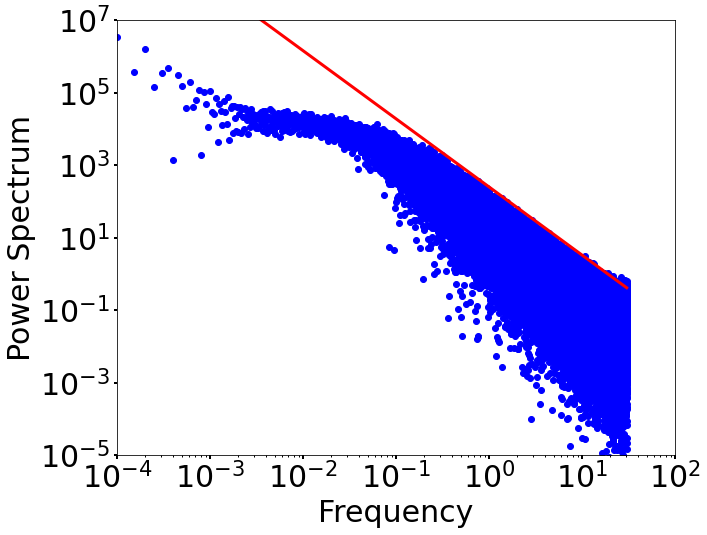}
  \end{minipage}
  \caption{
  \label{fig:spectrum_2022-09-23-19-00-00_L32_K0.0032_r0.02_regular_deg4_seed42}
  Power spectrum of the real-time trajectory of the fraction of cooperators in an iterated prisoner's dilemma game on a $4$-regular graph with cost-to-benefit ratio $r=0.02$ and temperature $T=3.2\times 10^{-3}$.
  The solid line is a power-law curve $S(f)\propto 1/f^{\alpha}$ with an exponent $\alpha=1.88$.
  }
\end{figure}

Figure \ref{fig:spectrum_2022-09-23-19-00-00_L32_K0.0032_r0.02_regular_deg4_seed42} shows the power spectrum at cost-to-benefit ratio $r=0.02$ and temperature $T=3.2\times 10^{-3}$.
We see two distinctive features similarly to those stated in relation to Fig. \ref{fig:spectrum_2022-09-17-22-14-13_L32_K1.0_r0.02_regular_deg4_seed42}.
One feature is the pink-noise behavior in the high-frequency region;
in fact, the upper envelope of the power spectrum above $f\gtrsim 0.5$ is well fitted by the power-law function $S(f)\propto 1/f^{\alpha}$ where the exponent is $\alpha =1.88$.
The other feature is that a few points over the frequency range between $f=10^{-4}$ and $f=10^{-3}$ have dominant spectral weights.
Thus, the long-term behavior in Fig. \ref{fig:2022-09-20-19-00-00_L32_K0.0032_r0.02_regular_deg4_seed42_linear_log} can be reproduced
by a small number of low-frequency modes.

\begin{figure}[htb]
  \begin{minipage}{0.99\linewidth}
    \centering
    \includegraphics[keepaspectratio, scale=0.25]{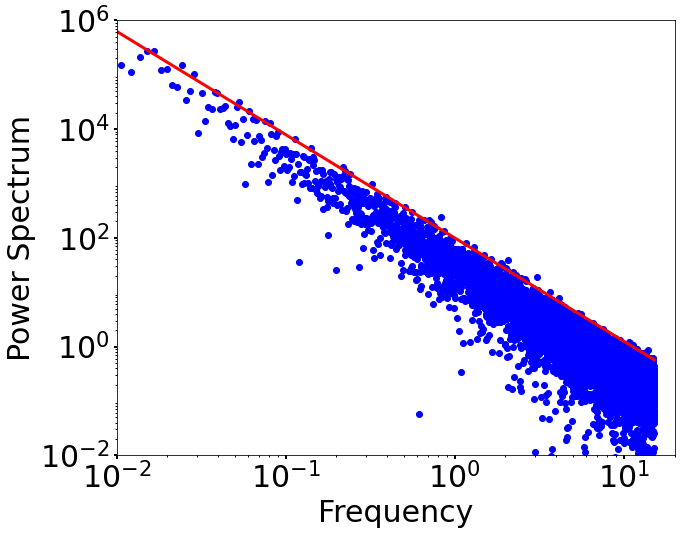}
  \end{minipage}
  \caption{
  \label{fig:spectrum_2022-09-20-20-38-18_L32_K100.0_r0.02_regular_deg4_seed42}
  Power spectrum of the real-time trajectory of the fraction of cooperators in an iterated prisoner's dilemma game on a $4$-regular graph with cost-to-benefit ratio $r=0.02$ and temperature $T=100.0$.
  The solid line is a power-law curve $S(f)\propto 1/f^{\alpha}$ with an exponent $\alpha=1.90$.
  }
\end{figure}

Figure \ref{fig:spectrum_2022-09-20-20-38-18_L32_K100.0_r0.02_regular_deg4_seed42} shows the power spectrum at cost-to-benefit ratio $r=0.02$ and temperature $T=100.0$.
Here we see that the pink noise extends over the entire frequency range.
The data below $f\lesssim 10^{-2}$ is absent because the fraction of cooperators either vanishes or saturates to 1 for time $t\gtrsim 10^2$.

\section{\label{sec:discussion}Discussion}

\begin{figure}[htb]
  \begin{minipage}{0.99\linewidth}
    \centering
    \includegraphics[keepaspectratio, scale=0.25]{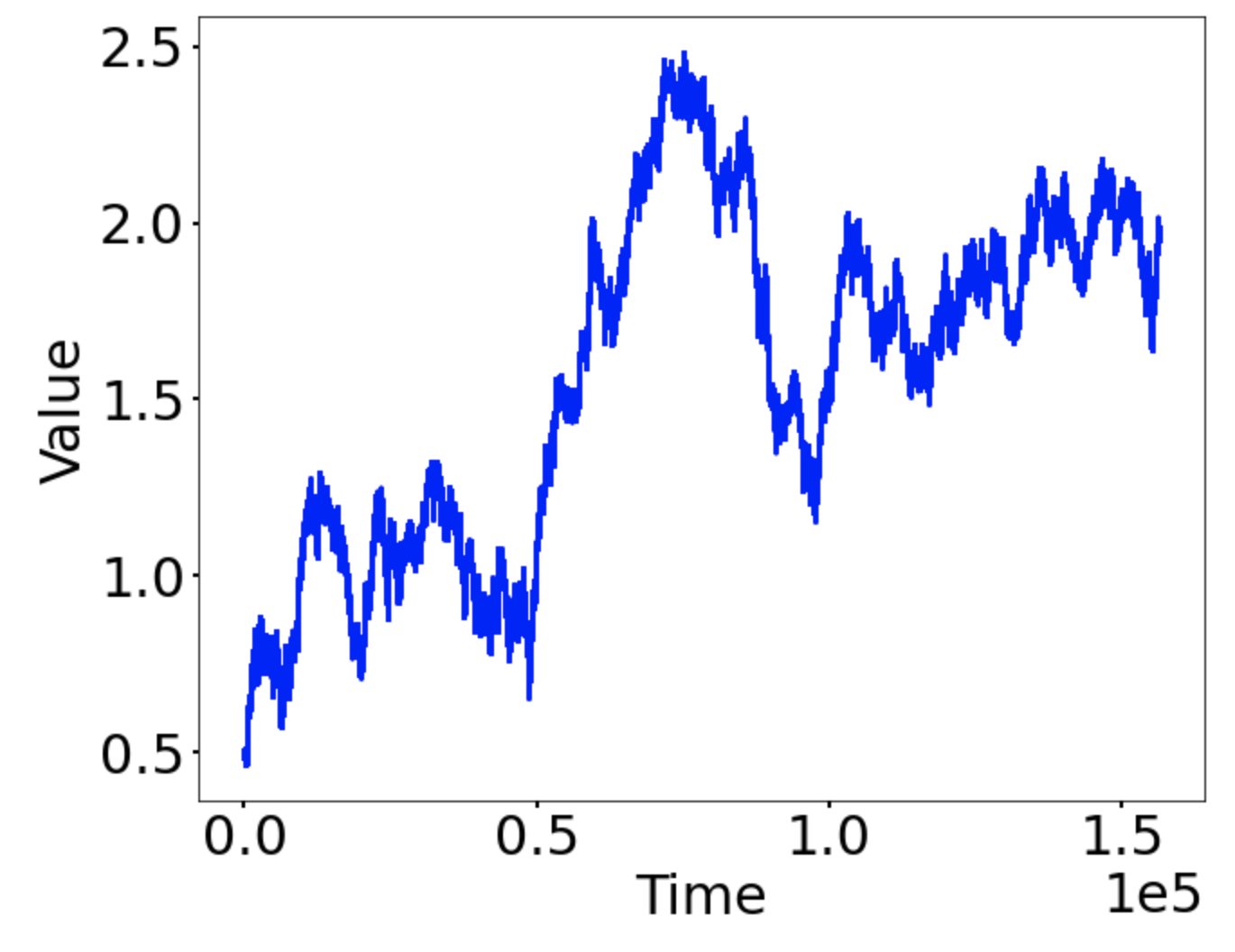}
  \end{minipage}
  \caption{
  \label{fig:random_walk}
  Real-time trajectory of the value obtained from a simple random walk.
  The value is initially set to be \(0.5\).
  For each step, the value increases by \(2^{-10}\), decreases by \(2^{-10}\) and is unchanged with uniform probability of \(1/3\).
  }
\end{figure}

\begin{figure}[htb]
  \begin{minipage}{0.99\linewidth}
    \centering
    \includegraphics[keepaspectratio, scale=0.25]{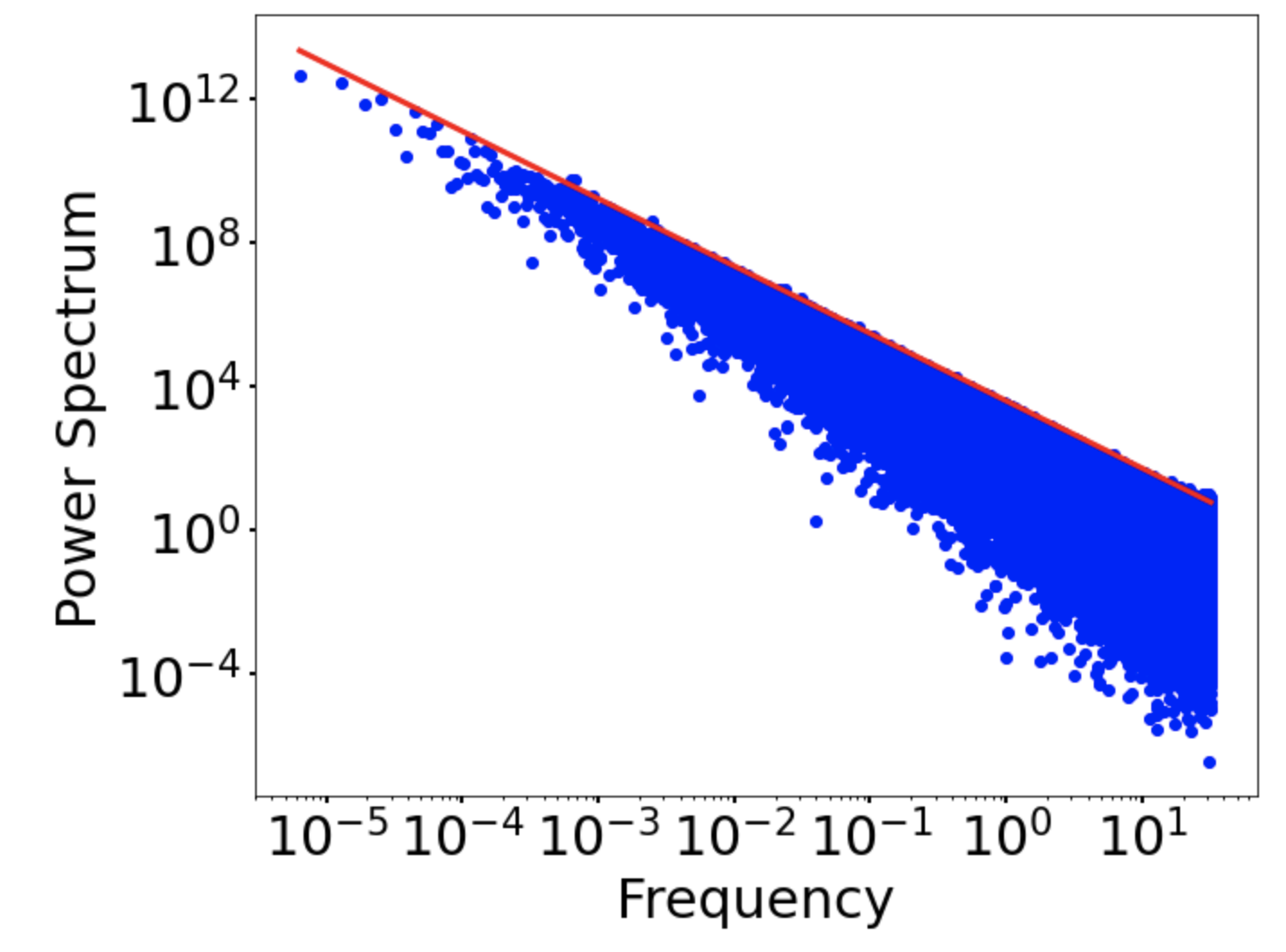}
  \end{minipage}
  \caption{
  \label{fig:spec_walk}
  Power spectrum of the real-time trajectory in Fig. \ref{fig:random_walk} of the value obtained from a simple random walk.
  The solid line is a power-law curve $S(f)\propto 1/f^{\alpha}$ with an exponent $\alpha=1.88$.
  }
\end{figure}

To investigate the origin of the pink-noise behavior in the power spectrum,
we perform a simple random walk.
More specifically, we consider a trajectory of a variable \(x\) that changes in time,
which is independent of the game-theoretical model we have considered.
The initial value of \(x\) is set to be \(x=0.5\), and at each time step, the value of \(x\) changes by
\(-2^{-10}, +2^{-10}\), and \(0\) with equal probability of \(1/3\).
Figure \ref{fig:random_walk} shows the trajectory of \(x\) that is generated by this simple random walk.
From the Fourier transform for the trajectory in Fig. \ref{fig:random_walk},
we obtain the power spectrum shown as blue dots in Fig. \ref{fig:spec_walk}.
The solid line in Fig. \ref{fig:spec_walk} is a power-law curve $S(f)\propto 1/f^{\alpha}$ with an exponent $\alpha=1.88$.
We find that this solid curve fits well with the upper part of a set of the blue dots.
This finding indicates that the simple random walk reproduces the pink-noise behavior,
similarly to the fraction of cooperators in the game-theoretical model we have considered.
Moreover, the power exponent of the simple random walk is close to that obtained in the game-theoretical model.

However, there is a notable difference between the simple random walk and the game theoretical model.
In Fig. \ref{fig:spec_walk}, which is the case of the simple random walk,
the spectrum fits well to the power-law curve over the entire range.
In the case of the game-theoretical model, meanwhile,
the low-frequency spectrum does not fits so well to the power-law curve,
and the spectrum flattens out in the range of \(f\lesssim 3\times10^{-3}\),
as in Fig. \ref{fig:spectrum_2022-09-20-19-00-00_L32_K0.1_r0.02_regular_deg4_seed42}.

\begin{figure}[htb]
  \begin{minipage}{0.99\linewidth}
    \centering
    \includegraphics[keepaspectratio, scale=0.25]{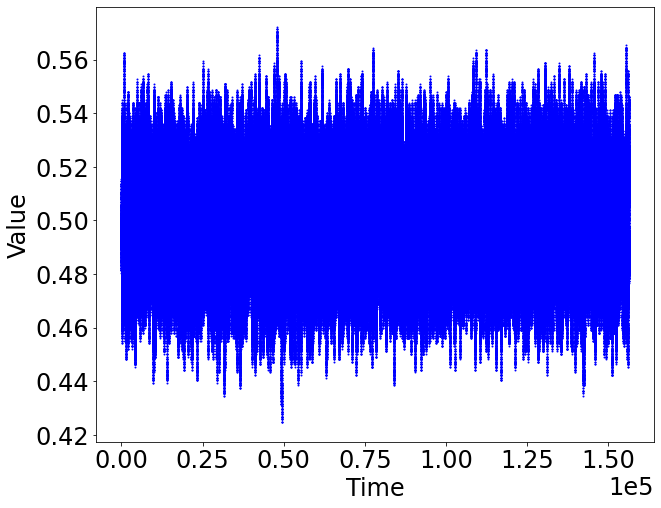}
  \end{minipage}
  \caption{
  \label{fig:trajectory_feedback20230623}
  Real-time trajectory of the value obtained from a modified random walk.
  The value is initially set to be \(0.50\).
  For each step, the value \(x\) increases by \(2^{-10}\), decreases by \(2^{-10}\) and is unchanged with probability of
  \(2x/3, 2(1-x)/3\) and \(1/3\), respectively.
  }
\end{figure}

\begin{figure}[htb]
  \begin{minipage}{0.99\linewidth}
    \centering
    \includegraphics[keepaspectratio, scale=0.25]{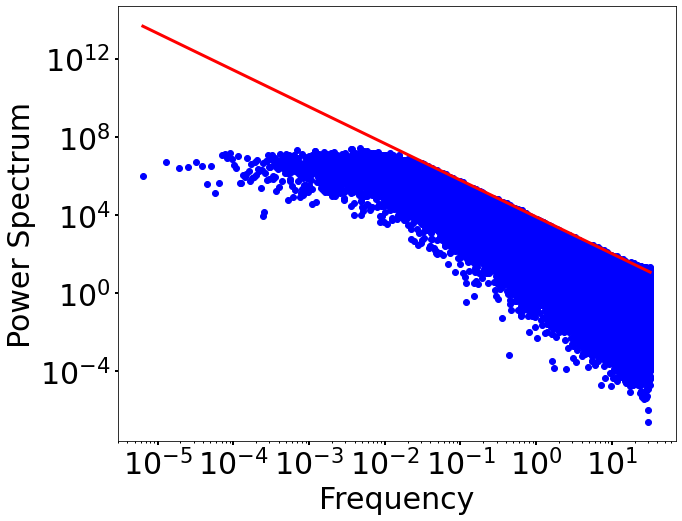}
  \end{minipage}
  \caption{
  \label{fig:spectrum_feedback_20230623}
  Power spectrum of the real-time trajectory in Fig. \ref{fig:trajectory_feedback20230623}
  of the value obtained from a modified random walk.
  The solid line is a power-law curve $S(f)\propto 1/f^{\alpha}$ with an exponent $\alpha=1.88$.
  }
\end{figure}

To understand the origin of the deviation of the spectrum from the fitting curve,
we perform a modified random walk.
More specifically, we consider a variable \(x\) whose transition probability depends on its own present value.
The value of \(x\) is initially set to be \(x=0.5\), and at each time step, the value of \(x\) changes by
\(-2^{-10}\), \(+2^{-10}\) and \(0\) with probability of \(2x/3, 2(1-x)/3\) and \(1/3\), respectively.
This setting suggests that \(x\) is always bounded by \(0\leq x\leq 1\) and that \(x\) is likely to decrease (increase)
when \(x\) is greater (smaller) than \(0.5\).
Figure \(\ref{fig:trajectory_feedback20230623}\) shows the trajectory of \(x\) that is generated by the modified random walk.
From the Fourier transform of the real-time trajectory in Fig. \ref{fig:trajectory_feedback20230623},
we obtain the power spectrum shown as blue dots in Fig. \ref{fig:spectrum_feedback_20230623}.
The solid line in Fig. \ref{fig:spectrum_feedback_20230623} is a power-law curve $S(f)\propto 1/f^{\alpha}$ with an exponent of $\alpha=1.88$.
In the high-frequency regime of \(f\gtrsim 2\times10^{-2}\), the solid line fits well with the upper envelope of the blue dots.
This means that the modified random walk reproduces the pink-noise behavior, which is also reproduced with the simple random walk.
In the low-frequency regime of \(f\lesssim 2\times 10^{-2}\), on the other hand,
we find that the blue dots deviate from the solid line and their upper envelope is almost flat.
This finding is consistent with the result obtained from the game-theoretical model
as in Fig. \ref{fig:spectrum_2022-09-20-19-00-00_L32_K0.1_r0.02_regular_deg4_seed42}.
Therefore, we find that the deviation from the pink noise in the game-theoretical model involves
the unbalanced transition probability of the fraction of cooperators.
In fact, at each time step, a cooperator (defector) is likely to be chosen to change his own action
when cooperators (defectors) dominates defectors (cooperators),
since a player is uniformly randomly chosen from all players.
Thus, the number of cooperators is likely to decrease (increase) when cooperators (defectors) are dominant
if temperature \(T\) is low enough to balance the transition probability.
As a result, the fraction of cooperators oscillates around one half for a long time
as shown in Fig. \ref{fig:2022-09-20-19-00-00_L32_K0.1_r0.02_regular_deg4_seed42_linear_log}.

For a high-temperature case, the payoffs of players
do not significantly contribute to the stochastic process in the game.
This is because the transition probability in Eq. (\ref{eq:pairwise-fermi}) is approximately one half
for a sufficiently high temperature \(T\), regardless of the difference \(p_x-p_y\) of payoffs.
Therefore, the fraction of cooperators for high temperature is expected to behave
similarly to the simple random walk.
We note that the fraction of cooperators is limited to values between zero and one,
but that the simple random walk has no such restriction.
Therefore, the trajectory of the fraction of cooperators can be regarded as
that of the simple random walk which is cut off by the boundaries of zero and one.
In fact, the trajectory in Fig. \ref{fig:2022-09-20-20-38-18_L32_K100.0_r0.02_regular_deg4_seed42}
of the fraction of cooperators with \(T=100\) does not significantly oscillate
before cooperators go extinct.
In addition, the shape of the spectrum in Fig. \ref{fig:spectrum_2022-09-20-20-38-18_L32_K100.0_r0.02_regular_deg4_seed42}
with \(T=100\) is similar to that of the simple random walk.
We note that with a large number \(n\) of players, we will obtain a long-time trajectory
and a low-frequency spectrum of the fraction of cooperators.
For a low-temperature case, in contrast, the game is significantly affected by the payoffs of players, and
the stochastic process in the game deviates in the low-frequency regime from the simple random walk.
In fact, the spectrum in Figs. \ref{fig:spectrum_2022-09-20-19-00-00_L32_K0.1_r0.02_regular_deg4_seed42} and \ref{fig:spectrum_2022-09-23-19-00-00_L32_K0.0032_r0.02_regular_deg4_seed42}
with \(T=0.1, 3.2\times10^{-3}\) exhibits pink-noise behavior for high frequency,
while the spectrum deviates from the fitting curve for low frequency.

For reference, we perform numerical simulation of the boundary game, which is also classified as the prisoner's dilemma game.
The boundary game is represented by TABLE \ref{table:boundary}.
\begin{table}[H]
  \caption{Payoff matrix of the boundary game.}
  \label{table:boundary}
  \centering
  \begin{tabular}{c|cc}
    Alice\textbackslash Bob & C & D\\ \hline
    C & $(1,1)$ & $(0,r)$\\
    D & $(r,0)$ & $(0,0)$
  \end{tabular}
\end{table}
The parameter \(r\) can be regarded as the strength of the temptation to choose defection.
For large \(r>1\), unilateral defection provides a greater payoff than mutual cooperation.
Then, mutual cooperation is not stable in a one-shot game. 
On the other hand, for small \(0<r<1\), mutual cooperation maximizes both players' payoffs.
Then, mutual cooperation is likely to occur.

\begin{figure}[htb]
  \begin{minipage}{0.99\hsize}
    \centering
    \includegraphics[width=7cm]{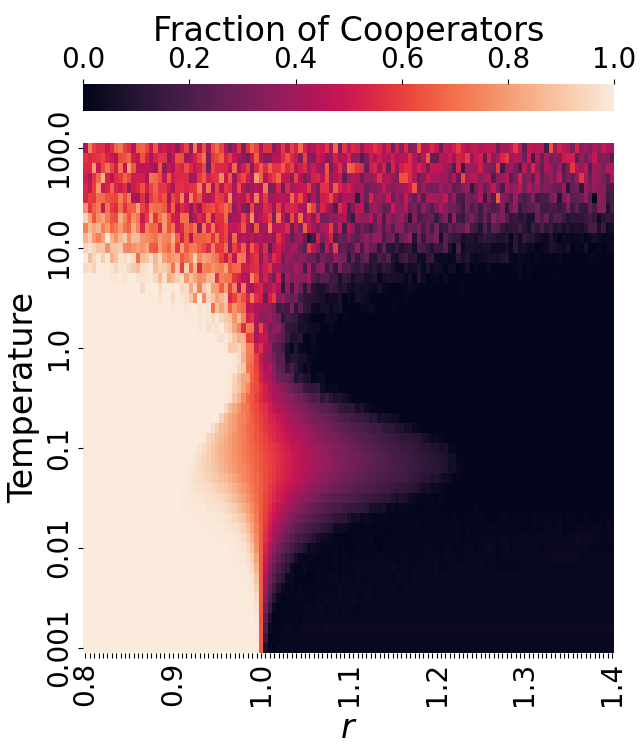}
  \end{minipage}
  \caption{
  \label{fig:heatmap_revised_fracdf_boundary_deg4_20231128_seed4}
  The fraction of cooperators appearing in the iterated boundary games on the 4-regular graph.
  }
\end{figure}

We similarly simulate an iterated boundary game on a \(4\)-regular graph, simply by replacing the game in TABLE \ref{table:one-parameter} by the boundary game in TABLE \ref{table:boundary}.
Figure \ref{fig:heatmap_revised_fracdf_boundary_deg4_20231128_seed4} shows the fraction of cooperators appearing in the iterated boundary game on the 4-regular graph.
The result appears qualitatively similar to the case of TABLE \ref{table:one-parameter} in the sense that the diamond-shaped mixed region exists.

\section{\label{sec:conclusion}Conclusion and Outlook}

In this paper, we have investigated the phase diagram of the iterated prisoner's dilemma game on a graph
as a function of the cost-to-benefit ratio and temperature.
We have also studied real-time dynamics of the fraction of cooperators in the game.
In Sec. \ref{sec:evolutionary_game}, we have formulated the iterated multiplayer game
where players shift their strategies according to the pairwise-Fermi update rule.
In Sec. \ref{sec:ensemble-averaged_square}, we have discussed the averaged quantities such as the fraction of cooperators and
the correlation of neighboring pairs in the game on a square lattice.
The phase-transition-like behavior can be seen in the game, and the correlation of neighboring pairs is found to be significant
in the mixed region, where cooperators and defectors coexist.
In Sec. \ref{sec:ensemble-averaged_regular}, we have examined the fraction of cooperators and the correlation of neighboring pairs
in the game on a regular graph.
We have found that the degree of a regular graph affects the boundaries of the mixed region,
and the diamond-shaped mixed region significantly shrinks for large degrees \(k\geq6\).
In Sec. \ref{sec:single}, we have studied the real-time trajectories of the fraction of cooperators of the game
for the cost-to-benefit ratio \(r=0.02\).
The power spectrum is found to behave as a pink noise in the sense that the upper envelope of the spectrum
can be fitted by a power-law curve over a wide frequency region.
In Sec. \ref{sec:discussion}, we have compared the dynamics of the game-theoretical model with the dynamics of the random-walk models.
As a result, the pink-noise behavior can be reproduced by the simple random walk.
Moreover, the deviation of the low-frequency spectrum from the pink-noise behavior in the game-theoretical model
is found to be reproduced by the modified random walk, where a negative-feedback effect is introduced.

We have discussed the phase-transition-like behavior found in the iterated game, which is analogous to phase transitions in physics
in the sense that several distinct phases of configuration appear depending on the two parameters, the cost-to-benefit ratio and temperature.
However, we have not given the strict definition of phase transitions in the game-theoretical model.
One possible definition may involve finite-size scaling analysis, which investigates
critical properties and scaling behavior of a finite-sized system.
Moreover, the averaged correlation of neighboring pairs in Eq. (\ref{eq:correlation_square}) is nonvanishing in the mixed region:
it is interesting to investigate how the correlation affects the real-time dynamics.

In this paper, we have focused on iterated prisoner's dilemma games on a square lattice and regular graphs; however, we can also consider games on other types of graphs
such as a triangular lattice, a hexagonal lattice and a scale-free network.
Especially, scale-free networks, where degrees \(k\) of nodes are distributed
as a power-law function proportional to \(\propto k^{-\gamma}\) with a scaling exponent \(\gamma\),
are widely observed in real world \cite{barabasi2003scale}.
Thus, iterated games on a scale-free network will be worth considering
in order to model real phenomena that arise from interactions among people.
In addition, it is known that in physics frustration arises in a triangular lattice model with antiferromagnetic interactions \cite{mezard1987spin}.
Considering this fact, we expect that a triangular lattice could give rise to a similar phenomenon in a certain iterated game.
One way to realize a frusrtated state in a game-theoretical model might be to adopt a payoff matrix
where payoffs are minimized by mutual cooperation and mutual defection and payoffs are maximized by disagreeing choices.
In this case, neighboring players will tend to choose distinct choices for higher payoffs.
However, it is impossible that three players on a triangle lattice choose distinct strategies,
since only two strategies can be chosen, and at least two pairs out of the three have to settle for the lowest payoffs.
Thus, frustration obtained in such a game-theoretical model will also be worth exploring.

\begin{acknowledgements}
  This work was supported by KAKENHI Grant No. JP22H01152 from the Japan Society for the Promotion of Science.
\end{acknowledgements}

\bibliography{apssamp}

\end{document}